\documentclass[11pt]{article}

\usepackage{graphicx,epsfig,amsthm,amssymb,amsbsy,amsmath,cleveref,enumitem,setspace}
\usepackage[title]{appendix}

\newcommand{\bb}[1]{\left({#1}\right)}					
\newcommand{\cc}[1]{\left\{#1\right\}}					
\newcommand{\op}[1]{\mathcal{#1}}
\newcommand{\ord}[1]{{\cal O}\bb{#1}}					
\newcommand{\ordx}[1]{{\cal O}}					
\newcommand{\abs}[1]{\left|#1\right|}					
\newcommand{\dabs}[1]{\abs{{#1}}}					
\newcommand{\dabss}[1]{\abs{{\mbox{\scriptsize$#1$}}}}					
		
\newcommand{\sfrac}[2]{\mbox{$\frac{#1}{#2}$}}	
\newcommand{\hf}{\mbox{$\frac12$}}
\newcommand{\tr}{{\sf Tr}}

\renewcommand{\v}[1]{{\bf #1}}

\newcommand{\BG}{$\op B$-$\op G$~}

\newcommand{\alphab}{{\mbox{\boldmath$\alpha$}}}

\newcommand{\eps}{\varepsilon}

\newtheorem{definition}{Definition}[section]

\crefname{equation}{}{}
\Crefname{equation}{}{}

\def\eps{\varepsilon}

\begin{document}

\title{Catastrophe conditions for vector fields in $\mathbb R^n$}
\author{Mike R. Jeffrey
\footnote{Department of Engineering Mathematics, University of Bristol, Ada Lovelace Building, Bristol BS8 1TW, UK, email: mike.jeffrey@bristol.ac.uk}
}
\date{\today}


\maketitle

\begin{abstract}
Practical conditions are given here for finding and classifying high codimension intersection points of $n$ hypersurfaces in $n$ dimensions. 
By interpreting those hypersurfaces as the nullclines of a vector field in $\mathbb R^n$, 
we broaden the concept of Thom's catastrophes to find bifurcation points of (non-gradient) vector fields of any dimension. We introduce a family of determinants $\op B_j$, such that a codimension $r$ bifurcation point is found by solving the system $\op B_1=...=\op B_r=0$, subject to certain non-degeneracy conditions. The determinants $\op B_j$ generalize the derivatives $\sfrac{\partial^j\;}{\partial x^j}F(x)$ that vanish at a catastrophe of a scalar function $F(x)$. 
We do not extend catastrophe theory or singularity theory themselves, but provide a means to apply them more readily to the multi-dimensional dynamical models that appear, for example, in the study of various engineered or living systems. For illustration we apply our conditions to locate butterfly and star catastrophes in a second order PDE. 
\end{abstract}


\newpage
\section{Introduction}

In this paper we give conditions that can be solved to locate degenerate critical points of a function $\v F:\mathbb R^n\times\mathbb R^p\to\mathbb R^n$. 
Letting $\v F=(F_1,F_2,...,F_n)$, a critical point in this context means a common zero of several functions 
\begin{align}\label{cp}
(F_1,F_2,...,F_n)=(0,0,...,0)\;,
\end{align}
in some variable $\v x\in\mathbb R^n$ with parameters $\alphab\in\mathbb R^p$. 
If the function $\v F$ is singular at a critical point $(\v x,\alphab)=(\v x_*,\alphab_*)$ then, under perturbation to any nearby $\alphab$, it will typically bifurcate to produce multiple critical points of $\v F$ in the space of $\v x$.

Quite simply, our aim here is to define a set of conditions 
\begin{align}\label{cond0}
\op B_1=\op B_2=...=\op B_r=0\neq\op G_r
\end{align}
whose solution, along with the condition $\v F=0$, provides the location of a codimension $r$ local bifurcation point in the vector field $\v F$. The functions $\op B_i$ are determinants of augmented Jacobian matrices that generalize the derivatives $\sfrac{d^i}{dx^i}$ which characterise catastrophes in one dimension (see e.g. \cite{ps96,t75,wp74}), while $\op G_r$ represents a set of non-degeneracy conditions. 

The main motivation behind this study comes from the problem of finding high codimension local bifurcation points of a vector field $\v F$, specifically to study steady states of an ordinary differential equation such as $\sfrac{d\;}{dt}\v x=\v F$, or of a partial differential equation such as $\sfrac{d\;}{dt}\v u+k\sfrac{\partial^2\;}{\partial x^2}\v u=\v F$. The results can also be used to characterise zeros of varieties $F_i$, or singular intersections of differentiable hypersurfaces defined by the equations $F_i=0$, for $i=1,...,n$. 


Given a map $\v F:\mathbb R^n\times\mathbb R^p\to\mathbb R^n$ and a particular point $(\v x_*,\alphab_*)\in\mathbb R^n\times\mathbb R^p$, singularity theory provides the means to classify any singularity that may occur there with great generality, see e.g. \cite{a93v,boardman67,mond20,montaldi21,takens74sing}. Our aim here is essentially to turn this around, to provide readily solvable conditions that can be solved to find the point $(\v x_*,\alphab_*)$ at which some suspected singularity or bifurcation occurs. To do so our subject matter must be significantly less general than singularity theory. We will look only for places where multiple critical points of a function $\v F$ coincide, and derive conditions that detect the {\it underlying catastrophe}, by which we will mean the coincidence of zeroes of the vector field, without reference to the vector field's directionality (and therefore without reference to phase portraits or stability in a dynamical system generated by $\v F$). 

It is notable, for instance, that modern numerical continuation packages (e.g. COCO, MatCont, AUTO \cite{coco,matcont,auto97}) do not yet contain a built-in condition to locate and continue general high codimension bifurcation points. Solvable conditions to locate such points are limited to low codimension cases, in particular the fold (or saddle-node) and cusp bifurcations (see e.g. \cite{k98}), or involving folds or cusps accompanied by more than one vanishing eigenvalue of the vector field's Jacobian (as in e.g. the `fold-Hopf' or `zero-Hopf' bifurcations \cite{a93v,gh02,k98}). The conditions \cref{cond0} are intended to provide functions that can be defined for general codimension $r$, and solved by continuing successive zeroes of the functions $\op B_1=0$, $\op B_2=0$, ... .

The classification of such points in either mappings or vector fields has particularly focused on showing equivalence to certain local forms. Arnold's classification in \cite{a93v}, for example, shows that the germ of a singularity in a system $\sfrac{d\;}{dt}\v x=\v F$ reduces to a small number of known classes, such as class $A_{\nu}$, which has ``one zero eigenvalue and $(\nu-1)$-fold degeneracy in the nonlinear terms'', and is locally reducible to a one dimensional normal form or `principal family' 
\begin{align}\label{prince}
\dot x=\pm x^{\nu+1}+\eps_\nu x^{\nu-1}+...+\eps_2x+\eps_1\;,
\end{align}
along a centre manifold through the bifurcation point. 

Families like \cref{prince} bear an obvious similarity to Thom's elementary catastrophes \cite{t75,thom55}, and one frequently encounters the recognisable signatures of such catastrophes in analysis of complex biological and engineering systems. Precisely locating the site of some suspected catastrophe or bifurcation, however, is nontrivial, and knowing that a vector field should be locally reducible to a form such as \cref{prince} does not provide a means to locate the variable $\v x$ and parameter $\alphab$ at which a given bifurcation occurs. 

\begin{figure}[h!]\centering\includegraphics[width=0.8\textwidth]{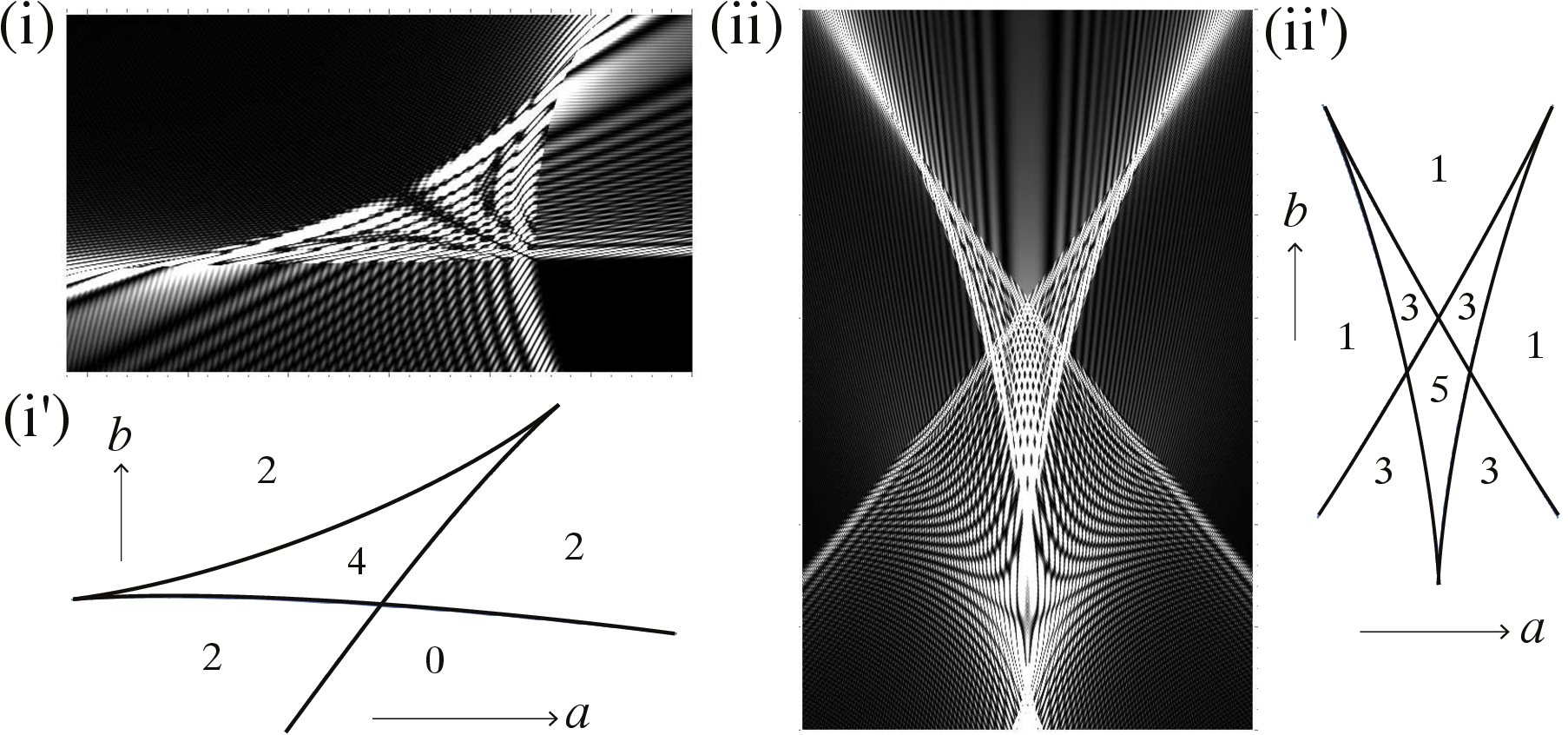}
\vspace{-0.3cm}\caption{\small\sf Two simulations of optical diffraction patterns near: (i) a swallowtail, (ii) a butterfly. The accompanying plots (i'-ii') are bifurcation diagrams of vector fields whose catastrophe structure mimics the diffraction images, using (i') $\v F(x,y)=\bb{a(1+x)+b(xy+\hf x-2)+x^2y+x^4+y-3x^2,y+2xy+x^2y}$, and (ii') $\v F(x,y)=\bb{a(x^2y^2+5)+bx-20x^3+x^5,y-x^3y+xy}$; numbers indicate the number of critical points of $\v F$ in each region. 
}\label{fig:diffraction}\end{figure}

\Cref{fig:diffraction} shows such signature patterns captured in the diffraction of light, 
in this case swallowtail and butterfly catastrophes (these are simulations, but actual photographs showing similar patterns with equal clarity can be found in Fig. 2.10 of \cite{berry89}). As the application of catastrophe theory grew in physics, numerous everyday examples like these were demonstrated in the scattering of light through water droplets, drinking vessels, and swimming pools, see \cite{berry77,berry89} for several examples. The bright fringes in \cref{fig:diffraction} are optical {\it caustics} --- fold catastrophes --- whose intersections produce cusps, swallowtails, and so on through successively higher {\it codimension}. The accompanying  $(a,b)$ parameter plane diagrams in \cref{fig:diffraction} show examples of the same catastrophe structure, but in the bifurcation structure a nonlinear vector field, found using the conditions we will introduce here.

The conditions $\op B_i=0$ given in \cref{cond0} reduce precisely to the one-dimensional derivatives $\sfrac{\partial^i\;}{\partial x_1^i}F_1=0$ for classes of vector fields that are equivalent to polynomial forms 
\begin{align}\label{F1}
\v F=\Big(x_1^{r+1}+\sum_{i=1}^{r}\alpha_{i}x_1^{r-i}\;,\;x_2\;,\;...\;,\;x_n\;\Big)\;,
\end{align}
which are consistent with applying Arnold's {\it principal families} \cref{prince} along some centre manifold --- the $x_1$ axis --- of an $n$-dimensional system. 
Typically, the bifurcation point ($\v x=\alphab=0$ in \cref{F1}) breaks up into up to $r+1$ critical points of $\v F$ under perturbations of the parameter $\alphab$, but to prove that the conditions $\op B_1=\op B_2=...=\op B_r=0\neq\op G_r$ imply this in general remains an open problem. 

It is important to distinguish between the {\it singularities} of a mapping $\v F$, and 
what we mean here by {\it catastrophes} of a vector field $\v F$. Singularities are points where a mapping $\v F$ is non-invertible, which involves derivatives of $\v F$, but not the value of $\v F$ itself. For a vector field $\v F$, we are often not concerned with singularities in this sense, but rather in points where zeros of $\v F$ encounter those singularities, forming local bifurcation points. These are what are typically referred to as {\it catastrophes} in dynamical systems theory, but we will use the term {\it underlying catastrophes} to distinguish them from standard uses of the term, and we exploit this property of them being `singular zeros' here to derive general conditions to locate and classify them. 

The pragmatic distinction we make between singularities and catastrophes is discussed in more detail in \cref{sec:nye}, using a particularly clear illustration due to {\it Nye et al} \cite{nye80,nye78}. 
We then outline the conditions and use them to define a family of underlying catastrophes in \cref{sec:cats}. A particular form of vector field is given in \cref{sec:primary}, for which the conditions reduce to something clearly consistent with one-dimensional catastrophes. Some examples for underlying catastrophes up to codimension 6 are outlined in \cref{sec:examples}, and some closing remarks are made in \cref{sec:conc}. Further illustrations are given for lower codimension underlying catastrophes in the Appendix, highlighting how the conditions work, along with some novel geometries.

\section{\!\!Catastrophes versus singularities for a vector field}\label{sec:nye}

What constitutes a singularity in a vector field was nicely illustrated by {\it Nye et al.} in their application to continuous media \cite{nye80,nye78}, using examples of sea ice flow and geostrophic winds. 

\begin{figure}[h!]\centering\includegraphics[width=0.8\textwidth]{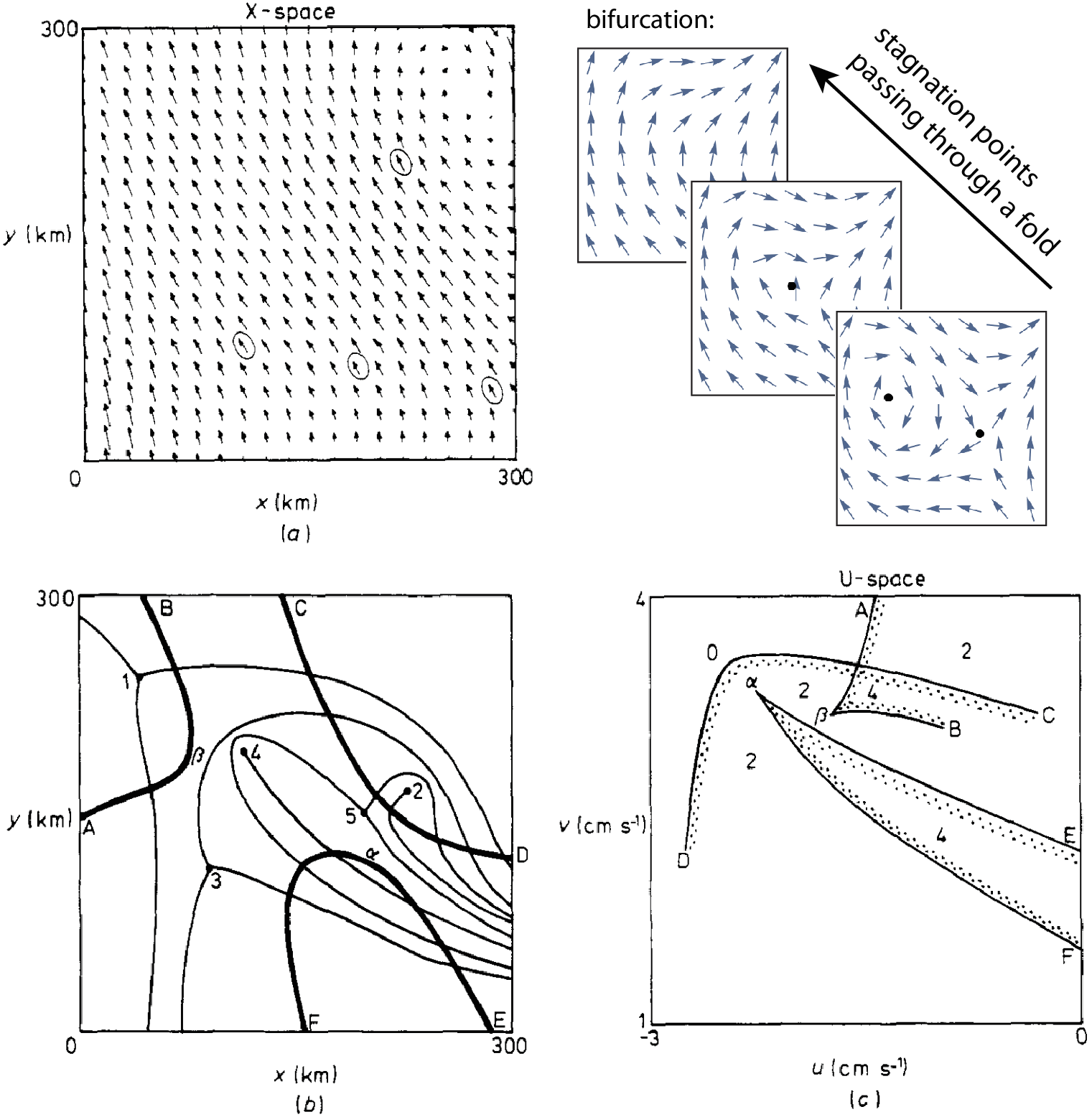}
\vspace{-0.3cm}\caption{\small\sf Singularities in a flow field of Arctic ice (a-c are reproduced from Figure 4 of \cite{nye78}). (a) The vector field $\v F=\bb{u(x,y),v(x,y)}$. (b) Singularities of the vector field: the bold curves (AB,CD,EF) are folds, where the Jacobian $J=\sfrac{d\v F}{d\v x}$ has determinant zero, the thin curves are trajectories of a vector $\v a$ that minimizes $|J\v a|$, and tangencies between these two sets of curves create cusp points $\alpha,\beta$. (The numbered points are (anti)umbilic points; for an explanation of these and the meaning of the thin curves see \cite{nye78}). Figure (b) is not easy to interpret, but the singularities are more apparent when shown in velocity space in (c), with numbers indicating how many pre-images $\v F$ has in regions bounded by the folds, e.g. the circled vectors in (a) show 4 pre-images of a typical point inside the region bounded by the curve EF. Top-right: sketch of stagnation points $\v F=0$ passing through a fold, adapted from Figure 5a of \cite{nye78}. 
}\label{fig:nyefield}\end{figure}

\Cref{fig:nyefield} illustrates the velocity field $\v F=(u,v)$ of such a medium in coordinates $\v x=(x,y)$. 
The vector field in \cref{fig:nyefield}(a) shows no obvious sign of the singularities, whose loci are plotted in (b), in particular the folds (the bold curves AB,CD,EF) and cusps (the points $\alpha,\beta$). The meaning of these singularities is evident only when plotted in velocity space in (c), where we see that folds delimit regions on which the pre-images of the vector field change, and these curves of folds have non-differentiable corners at the cusps. 

The singularities are relevant to continuous media as they manifest in the strain of a medium, which involves derivatives of the flow field. In applications outside continuum mechanics, however, the notion of strain often has no practical meaning. To {\it attempt} to visualise the singularity, {\it Nye et al.} also describe how one may take a point from \cref{fig:nyefield}(c) and locate all pre-images of it in (a). For example, the four points circled in (a) map to a single point in the region bounded by the EF fold in (c). As parameters vary, these four vectors coalesce and annihilate in pairs as they pass over the fold curves in (b), constituting pre-images of the field disappearing pairwise as a point $(u,v)$ traverses the fold curves in (c). This is not easy to observe, of course, because it does not constitute a topological change in the vector field $\v F$ itself. 

Things are different if we consider points where $\v F=0$, i.e. critical points. When critical points of $\v F$ pass through a fold, the effect is clearly visible as a topological change in the vector field. The top-right of \cref{fig:nyefield}, for example (also loosely adapted from Figure 5a of \cite{nye78}), shows two critical points born through a saddle-node bifurcation as they pass through a fold. 
This topological change --- the appearance or disappearance of critical points of $\v F$ --- is what we will term here an {\it underlying catastrophe}. The `underlying' signifies that we care only about the vanishing of $\v F$, not the directions of $\v F$ nearby, allowing us to extend the usual notion of a `catastrophe' beyond scalar or gradient fields. 

In studying critical points of $\v F$, we have access to something more immediate than the degeneracies of the Jacobian $J=\sfrac{d\v F}{d\v x}$, namely the vanishing of the vector field components $\v F=(F_1,...,F_n)$. Each higher codimension of catastrophe can be identified as occurring where successive singularity sets, which we will define by conditions $\op B_i=0$, become tangent to each other. (In this aspect our scheme is similar in spirit to Whitney's, recalling from the caption of \cref{fig:nyefield} that the cusps in (b) arise where the bold and fine curves are tangent.) 

Let us now introduce these conditions and the geometry behind them.

\section{The catastrophe conditions}\label{sec:cats}

Consider a set of functions 
\begin{align}
F_i=F_i(x_1,...,x_n;\alpha_1,...,\alpha_p)\qquad {\rm for} \qquad i=1,...,n,\;
\end{align} 
where $x_i$ are variables and $\alpha_i$ are parameters. In vector form we write $\v F=\v F(\v x,\alphab)$, with $\v x=(x_1,...,x_n)$, $\alphab=(\alpha_1,...,\alpha_p)$, and $\v F=(F_1,...,F_n)$. Let $\v F:U\times A\to V$ be a $\op C^k$ differentiable function in the variable $\v x\in U$, for a parameter $\alphab\in A$, with $U,V\subset\mathbb R^n$, $A\subset\mathbb R^p$. We assume $k>r$ and $p\ge r$, where $r$ will be the codimension of the underlying catastrophe introduced in \cref{def:sings} below. 

For convenience we will typically just write `$\v F$' for the vector field, rather than making the dependence on `($\v x,\alphab)$' explicit. 

We wish to characterise points where $\v F=0$. 

\begin{definition}\label{def:zero}
A {\bf critical point} of $\v F$ is a point $(\v x_*,\alphab_*)\in U\times A$ that satisfies 
\begin{align}\label{singr}
&F_1=...=F_n=0\;.
\end{align}
\end{definition}

A bifurcation occurs where a critical point is degenerate. This happens at an intersection of the hypersurfaces $F_i=0$ where the Jacobian matrix $\frac{d\v F}{d\v x}$ has rank less than $n$, so its determinant $\abs{\sfrac{d\v F}{d\v x}}$ vanishes. 
This also means that the gradient vectors $\sfrac{d\;}{d\v x}F_1,...,\sfrac{d\;}{d\v x}F_n$, are linearly dependent, but we shall assume in general that if we take any $n-1$ of these gradients, they {\it are} linearly independent. 
We then need a notion of {\it how} degenerate the point is, in some sense 
the order of contact $r$ between the hypersurfaces, determining how many critical points may exist locally as we vary $\alphab$. 
We will first define conditions for this purpose in \cref{sec:BG}, before describing where they come from in \cref{sec:geometry}.

\subsection{The conditions}\label{sec:BG}

Let us first define determinants $\op B_i$, for $i=2,...,r,$ as 
		\begin{align}\label{Br}
		\op B_{i}&=\abs{\frac{\partial(\op B_{i-1},F_2,...,F_{n})}{\partial(x_1,...,x_n)}}\;,\qquad  
	\op B_1=\abs{\frac{\partial(F_1,...,F_{n})}{\partial(x_1,...,x_n)}}\;,
		\end{align}
		hence $\op B_1$ is just the determinant of the Jacobian of $\v F$, while each $\op B_i$ is a determinant in which the previous $\op B_{i-1}$ replaces the vector field component $F_1$. 
		These will be used for detecting singular intersections of the hypersurfaces $F_i=0$. 
		
		To establish non-degeneracy requires 
		a larger family of functions, firstly the extended determinant
		\begin{align}\label{BrGr0}
		\op G_1=\abs{\frac{\partial(F_1,...,F_{n},\op B_{1})}{\partial(x_1,...,x_n,\alpha_1)}}\;,
		\end{align}
		and its extensions to higher codimensions. To define these we need a little notation. For any scalar $V$, let us denote $$(F_1,...,F_n)\backslash^kV=(F_1,...,F_{k-1},V,F_{k+1},...,F_n)\;,$$ 
		so for instance the numerator of $\op B_i$ in \cref{Br} can be written $\partial(F_1,...,F_n\backslash^1 \op B_{i-1})$. 
		Let $I(s)$ denote an $s$-length string $I(s)=i_1...i_{s}$ of symbols $i_j\in\cc{1,...,n}$, and define a generalization of the determinants $\op B_i$ as
		\begin{align}\label{Bri}
		\op B_{s,I(s-1)}&=\abs{\frac{\partial(F_1,...,F_n)\backslash^{i_{s-1}}\op B_{s-1,I(s-2)}}{\partial(x_1,...,x_n)}}
		\end{align}
		for $s=2,...,r,$ defining $\op B_{1,I(0)}=\op B_1$. Note that $\op B_{i,1...1}\equiv\op B_i$. Lastly then, we can define a set of extended determinants of these functions given by
		\begin{align}\label{Gri}
		\op G_{r,I(r-1)}&=\abs{\frac{\partial(F_1,...,F_{n},\op B_{1},\op B_{2,I(1)},...,\op B_{r,I(r-1)})}{\partial(x_1,...,x_n,\alpha_1,...,\alpha_{r})}}\;,
		\end{align}
		defining $\op G_{1,I(0)}=\op G_1$. 
		 For ease of reference these `\BG functions' are expanded in \cref{sec:BGexpanded} for the first few $r$, and we will look at example calculations in \cref{sec:examples} and \cref{sec:further}.

			If the vector field $\v F$ depends on more than $r$ parameters (i.e. $p>r$), then we may denote that we are considering variation with respect to some subset of them in a superscript as $\op G_{r,I(r-1)}^{\alpha_1...\alpha_r}$. 
		
		Looking slightly ahead to the role of these $\op B$ and $\op G$ functions, to locate a given catastrophe, we are only required to solve the conditions $\op B_i=0$ for some $i=1,...,r$. This is typically a tractable problem, either analytically or numerically. 
		Notably, the family of functions $\op G_{r,i_1...i_{r-1}}$ defined by \cref{Gri} is rather larger, namely $n^{r-1}$ functions for a given $r$, but one only needs evaluate these to check they are non-vanishing. It is not known at present whether all of the functions $\op G_{r,i_1...i_{r-1}}$ are independent or some are equivalent, certainly in some cases many of them become trivial, but possible simplification of these conditions are left to future study. 

\medskip
	
Using these we propose the following definitions.

\begin{definition}\label{def:sings}
A vector field $\v F:U\times A\to V$ exhibits an {\bf underlying catastrophe} of codimension $r$ if it has a critical point $(\v x_*,\alphab_*)\in U\times A$ at which  
\begin{align}\label{singr}
&\op B_1=...=\op B_{r}=0\neq\op B_{r+1}\;.
\end{align}
We say the underlying catastrophe is {\bf full} if the non-degeneracy conditions
	\begin{align}\label{singdeg}
	\op G_{r,i_1...i_{r-1}}\neq0\;,
	\end{align}
	hold at $(\v x_*,\alphab_*)$ for every $i_j\in\cc{1,...,n}$. 
\end{definition}

The conditions \cref{singr}-\cref{singdeg} define a codimension $r$ underlying catastrophe as occurring where the contact between the hypersurfaces $F_i=0$, $i=1,...,n$, is at least of order $r$, in the sense that the augmented Jacobian determinants $\op B_i$ vanish for orders $i=1,..,r$. 
Following Thom's classification in $n=1$ dimensions (see e.g. \cite{ps96,t75,wp74}), we refer to these underlying catastrophes as the {\it fold} ($r=1$), {\it cusp} ($r=2$), {\it swallowtail} ($r=3$), {\it butterfly} ($r=4$), {\it wigwam} ($r=5$), {\it star} ($r=6$), etc..

We propose this as a working definition. Its purpose is to identify the location of some catastrophe that underlies some singularity or bifurcation, whose proper equivalence class can then be found by standard local analysis. 
By giving up the strict adherence to equivalence classes in this way, we obtain a practical method to locate bifurcation points via their underlying catastrophe only. 

The name {\it underlying catastrophe} highlights that we are extending the use of the term `catastrophe' beyond functions that are reducible to Thom's elementary catastrophes, but whose zeroes bifurcate in an analogous way. Moreover in \cref{sec:primary} we provide classes for which these are directly related to Thom's catastrophes. An alternative name might be `{\it $n$-catastrophes}'. In any event, this terminology is only preliminary, since the proper classification of these points is known and can be established using singularity theory such as \cite{a93v}. The purpose of \cref{def:sings} is merely to provide a method to locate those points in the first place. 
One hopes that these definitions can be refined in the context of singularity theory, by relating the conditions \cref{singr}-\cref{singdeg} rigorously to ranks and ideals of germs, in future work. 

An important notion here is of being `full' in \cref{def:sings}, which ensures that the system \cref{singr} is uniquely solvable, but should not be mistaken as the definition of an equivalence class. 
In some circumstances a vector field that is not `full' can be made so by the removal of redundant dimensions. For example, the dynamical system $(\dot x,\dot y)=(ky+a+bx+x^3,y)$ has a cusp at $x=y=a=b=0$, but is only `full' in $\mathbb R^2$ provided $k\neq0$, so that $\sfrac{d\;}{d\v x}F_1=(0,k)$ is not identically zero at point of interest. If $k=0$ then instead we can use the same definitions but must restrict the problem to the $x$-dimension only, as then the cusp in $\dot x=a+bx+x^3$ is again `full'. We explore these issues in further examples in \cref{sec:further}. 

More precisely, the conditions $\op G_{r,I}\neq0$ ensure that where the sets $\op B_i=0$ for $i=1,...,r$, intersect, their gradient vectors are linearly independent, without which the conditions \cref{singr} may not be solvable. 
Note that it does not matter in \cref{singr} whether we evaluate each $\op B_i$, or any of the permutations $\op B_{i,I(i-1)}$, as generically, at a point where $\op B_1=…=\op B_r=0$, a family of $n$ linearly independent vectors cannot be formed from the gradient vectors to the sets $F_1=0$, ... $F_n=0$, or any of the sets $\op B_i=0$ for $i<r$. So we are guaranteed that if one choice of $\op B_{i,I(i-1)}$ vanishes for some string $I(i-1)$, they vanish for any string $I(i-1)$, hence in defining $\op B_i$ we are free to make the choice $\op B_i\equiv\op B_{i,1...1}$ in \cref{Br}. The non-vanishing of the $\op G_{r,I}$ ensures that none of the $\op B_{i\le r}$ vanish trivially, so this choice of $\op B_i$ is without loss of generality.

\subsection{Geometric interpretation}\label{sec:geometry}

The idea behind \cref{def:sings} is that catastrophes occur where the intersection of the hypersurfaces $F_i=0$ has an $r^{th}$ order contact at $(\v x_*,\alphab_*)$, such that in a neighbourhood of $\alphab_*$ the system has up to and including $r+1$ critical points. 

In $n=1$ dimension, the critical points of $F_1$ can be identified with turning points of Thom's potential functions \cite{ps96,t75}. In a codimension $r$ catastrophe, $r+1$ critical points of $F_1$ coincide at a point $c$ where $\sfrac{d\;}{dx_1}F_1=...=\sfrac{d^r\;}{dx_1^r}F_1=0$. 
Each successive $\sfrac{d^i\;}{dx_1^i}F_1$ that vanishes implies at least an $i^{th}$ order degeneracy of the given root of $F_i=0$ at $c$. 

The purpose of each function $\op B_i$ is to generalize these derivatives, by thinking of a fold being a degeneracy of zeros of $F_1$ due to coinciding with the set $\sfrac{d\;}{dx_1}F_1=0$, a cusp being a degeneracy between the fold and the zeros of $F_1$ due to coinciding with the set $\sfrac{d^2\;}{dx_1^2}F_1=0$, and so on. We can then generalize those conditions to a vector field $\v F$. 
In two dimensions, for example, the zero sets of the two functions $F_1=x_1^{p}-x_2$ and $F_2=x_2$ clearly have order $p$ contact at the origin, as the distance between them grows as $|x_1|^p$. For this we have $\op B_i=\sfrac{d^i\;}{dx_1^i}F_1=\ord{|x_1|^{p-i}}$. Under perturbation this perturbs into up to $p$ critical points. 

Now take any $n>1$. Consistent with singularity theory, the mapping $\v F:\mathbb R^n\times\mathbb R^p\to\mathbb R^n$ is non-singular at a point $c$ if $\op B_1\neq0$, since $\op B_1$ is simply the determinant of the Jacobian of $\v F$. Letting this $c$ be a point where $\v F=0$, this critical point is therefore non-degenerate. 

\begin{figure}[h!]\centering\includegraphics[width=0.85\textwidth]{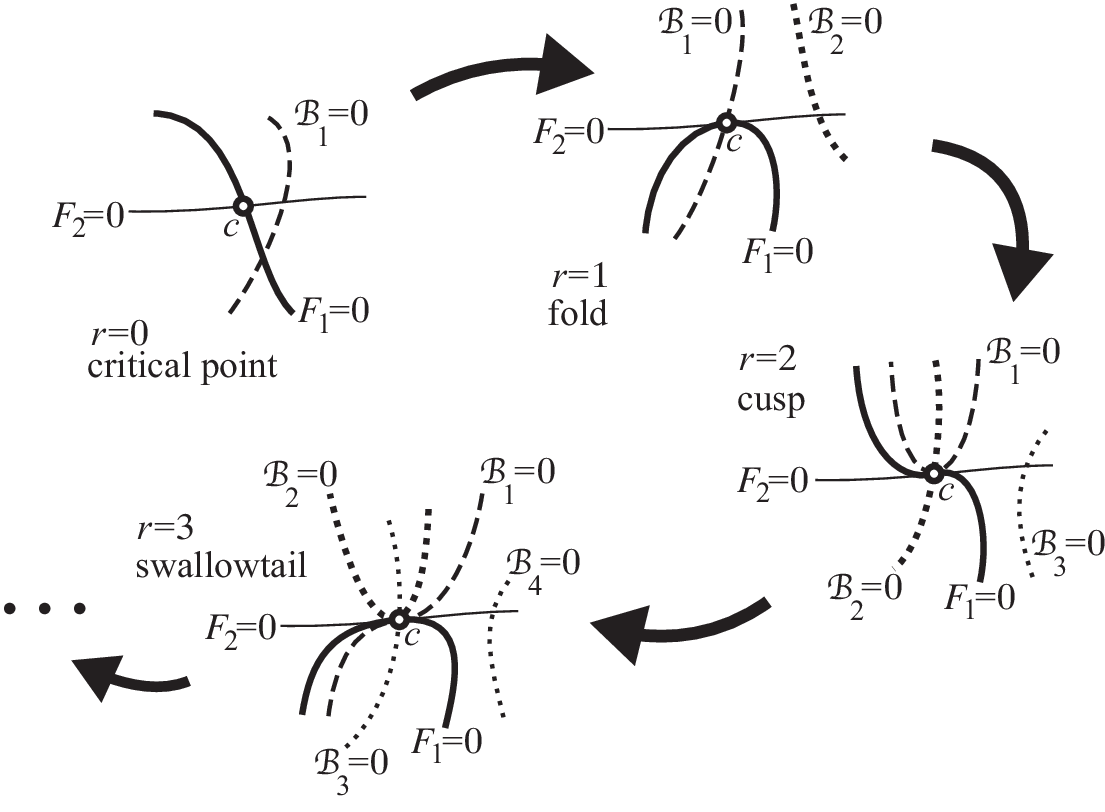}
\vspace{-0.3cm}\caption{\small\sf Illustration of the catastrophe definitions in two dimensions, writing $F=(f,g)$ for convenience. From top left: when $r=0$, $f=0$ and $g=0$ intersect transversally at $c$ to form a critical point, and the curves $\op B_r=0$ lie elsewhere; when $r=1$, $\op B_1=0$ intersects $c$ so $f=0$ and $g=0$ have quadratic contact; when $r=2$, $\op B_2=0$ intersects $c$ so $f=0$ and $g=0$ have cubic contact; when $r=3$, $\op B_3=0$ intersects $c$ so $f=0$ and $g=0$ have quartic contact (and in each case the other curves $\op B_{s>r}$ lie away from $c$); etc.. 
}\label{fig:trans}\end{figure}

\Cref{fig:trans} illustrates this non-degenerate ($r=0$) critical point in two dimensions, along with the codimensions $r\ge1$ following the argument below. 

If $\v F=\op B_1=0$ at $c$ then the critical point of $\v F$ lies at an underlying catastrophe. We have then to characterise its codimension. 

Generically, the gradient vectors $\sfrac{\partial\;}{\partial\v x}\op B_1$, and any $n-1$ of the gradient vectors $\sfrac{\partial\;}{\partial\v x} F_i$, will be linearly independent at $c$ (case $r=1$ in \cref{fig:trans}). Let $\v F$ depend on some parameter $\alpha_1$ such that $\alpha_1=0$ at $c$. If the gradient vectors $\sfrac{\partial\qquad}{\partial(\v x,\alpha_1)}\op B_1$ and $\sfrac{\partial\qquad}{\partial(\v x,\alpha_1)}F_i$ 
are linearly independent at $c$, then
		\begin{align}
		\op G_1&=\abs{\frac{\partial(F_1,...,F_{n},\op B_1)}{\partial(x_1,...,x_n,\alpha_1)}}\neq0\;
		\end{align}
at $c$. 
This then defines an {\it underlying fold catastrophe} of the vector field $\v F$ (in the sense of \cref{def:sings}), which unfolds a bifurcation of two critical points of $\v F$ as $\alpha_1$ varies through zero. 

If the fold catastrophe set $\op B_1=0$ lies tangent to the hypersurfaces $F_i=0$ at $c$ in the space of $\v x\in U$ (case $r=2$ in \cref{fig:trans}), 
then since the gradient vectors $\sfrac{\partial\;}{\partial\v x} F_i$ are themselves linearly dependent at $c$ (since $\op B_1=|\sfrac{\partial\;}{\partial\v x}\v F|=0$), then $\sfrac{\partial\;}{\partial\v x}\op B_1$ lies in the space spanned by {\it any} choice of $n-1$ of the normals $\sfrac{\partial\;}{\partial\v x} F_i$, hence all of the determinants 
		\begin{align}
		\op B_{2,i}&=\abs{\frac{\partial(F_1,...,F_{i-1},\op B_1,F_{i+1},...,F_n)}{\partial(x_1,...,x_n)}}\;,
		\end{align}
will be zero for any $i\in\cc{1,...,n}$. We only need to evaluate one of these $\op B_{2,i}$ in \cref{Br}, so we make the choice to define $\op B_2=\op B_{2,1}$. If $\v F$ depends on some parameters $(\alpha_1,\alpha_2)$ such that $\alpha_1=\alpha_2=0$ at $c$, and the normal vectors $\sfrac{\partial\qquad}{\partial(\v x,\alpha_1,\alpha_2)}\op B_2$, $\sfrac{\partial\qquad}{\partial(\v x,\alpha_1,\alpha_2)}\op B_1$, and $\sfrac{\partial\qquad}{\partial(\v x,\alpha_1,\alpha_2)}F_i$, are all linearly independent at $c$
, then
		\begin{align}
		\op G_{2,i}&=\abs{\frac{\partial(F_1,...,F_{n},\op B_1,\op B_{2,i})}{\partial(x_1,...,x_n,\alpha_1,\alpha_2)}}\neq0\;
		\end{align}
at $c$, for any $i=1,...,n$. 
This then defines an {\it underlying cusp catastrophe} of the vector field $\v F$ (again in the sense of \cref{def:sings}), which unfolds a bifurcation of three critical points of $\v F$ as $\alpha_1$ and $\alpha_2$ vary through zero. 

We proceed in a similar manner through successive codimensions of catastrophe. If the cusp catastrophe set $\op B_2=0$ lies tangent to the hypersurfaces $F_i=0$ at $c$ in the space of $\v x\in U$, we obtain an underlying swallowtail catastrophe, and so on through $r=3,4,5,...$.

\begin{figure}[h!]\centering\includegraphics[width=0.75\textwidth]{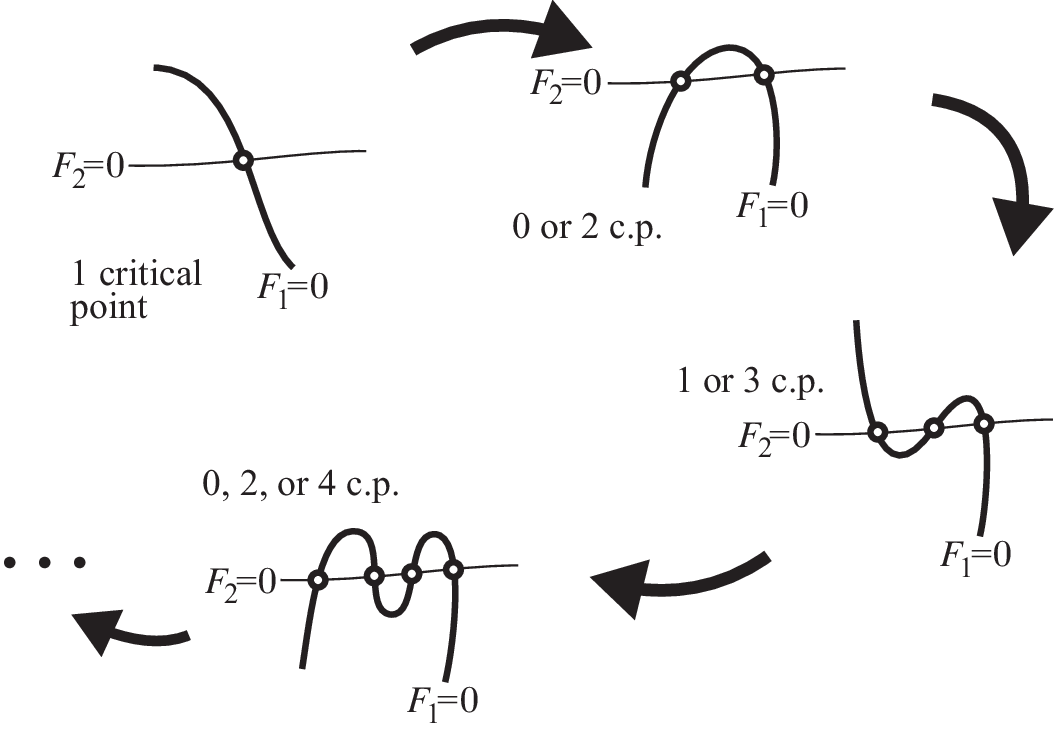}
\vspace{-0.3cm}\caption{\small\sf A typical perturbation of \cref{fig:trans} results in $r+1-2i$ critical points (c.p.) for $0\le i\le(r+1)/2$ (only the maximum $r+1$ is shown in the picture). 
}\label{fig:pert}\end{figure}

\Cref{fig:pert} illustrates how a perturbation of \cref{fig:trans} to nearby parameter values results in up to $r+1$ critical points. 
%

\section{Primary forms}\label{sec:primary}

It is not in our interest here to define normal forms for the catastrophes or bifurcations of vector fields, since these are already provided in \cite{a93v}. But it is instructive to embed the one-dimensional principal families \cref{prince} in an $n$-dimensional vector field, and we shall see how the functions $\op B_r$ then reduce in a natural way. 

Let us define {\it primary forms} having the general expression
\begin{align}\label{primary}
\v F=\Big(\;\phi(x_1)+\underline{k}\cdot\underline{x}\;,\;\lambda_2x_2\;,\;...\;,\;\lambda_nx_n\;\Big)
\end{align}
where $\underline{k}\cdot\underline{x}=k_2x_2+...+k_nx_n$ (the underline denoting that these are $(n-1)$-dimensional vectors), and 
\begin{align}
\phi(x_1)&=x_1^{r+1}+\sum_{i=1}^{r}\alpha_{i}x_1^{i-1}\;,\\
	&=x_1^{r+1}+\alpha_{r}x_1^{r-1}+...+\alpha_{2}x_1+\alpha_{1}\;.\nonumber
\end{align}
These are consistent with Arnold's {\it principal families} \cite{a93v} embedded along the $x_1$-axis of a system with dimension $n>1$. 
The terms $\lambda_2,...,\lambda_n$, are just eigenvalues of the Jacobian of $\v F$ associated with eigenvectors normal to the $x_1$ axis, and generically are nonzero. The terms $k_2,...,k_n,$ are nonzero constants that ensure the system \cref{primary} is {\it full} as per \cref{def:sings}, and that although the gradients $\sfrac{d\;}{d\v x}F_1,...,\sfrac{d\;}{d\v x}F_n$, are linearly dependent at the catastrophe point, any $n-1$ of these gradients are typically {\it are} linearly independent. The terms $\alpha_1,...,\alpha_r,$ are bifurcation parameters. 

For the system \cref{primary} the underlying catastrophe conditions take the simple form of derivatives in $x_1$. 
To show this, define an $(n-1)$-dimensional square matrix $\underline{\underline{\Lambda}}=\operatorname{diag}(\lambda_2,...,\lambda_n)$, such that $|\underline{\underline{\Lambda}}|=\lambda_2...\lambda_n$. Let $\underline0$ represent the $(n-1)$-dimensional zero row or column vector as appropriate, and denote the derivatives of $\phi$ with respect to $x_1$ as $\phi'$, $\phi''$, etc.. 
Then by straightforward calculation,
		\begin{align}\label{mB1}
		\op B_1&=\abs{\sfrac{\partial(F_1,...,F_{n})}{\partial(x_1,...,x_n)}}
		=\dabs{\begin{array}{cccccc}
		\phi'(x_1)	&\underline{k}\\
		\underline0	&\underline{\underline{\Lambda}}
		\end{array}}
		=\lambda_2...\lambda_n\;\phi'(x_1)\;,
		\end{align}
		and for $r>1$, 
		\begin{align}\label{mB2}
		\op B_{r}&=\abs{\sfrac{\partial(\op B_1,F_2,...,F_n)}{\partial(x_1,...,x_n)}}
		=\dabs{\begin{array}{cccccc}
		|\underline{\underline{\Lambda}}|^{r-1}\phi^{(r)}(x_1)	&\underline0\\
		\underline0	&\underline{\underline{\Lambda}}
		\end{array}}\qquad\quad\nonumber\\&
		=(\lambda_2...\lambda_n)^r\phi^{(r)}(x_1)\;.
		\end{align}	
In addition, clearly $\phi'(x_1)$ is also an eigenvalue of the Jacobian, some $\lambda_1=\phi'(x_1)$, hence $\op B_1=\lambda_1\lambda_2...\lambda_r$ is just the determinant of the Jacobian of $\v F$. 

We can make a connection between these expressions and singularity theory in different ways. Perhaps the most simple is to note that the function $\phi(x_1)$ is the normal form of a singularity of the mapping $$(x_1,\alpha_1,...,\alpha_r)\mapsto\bb{\phi(x_1),\alpha_1,...,\alpha_r}\;,$$ called a {\it Morin singularity}, see e.g. \cite{fukuda88,mond20,morin65}). 
The family of vector fields $\v F$ are unfoldings of 
\begin{align}\label{primaryunfold}
\v F_{0}=\Big(\;x_1^{r+1}+\underline{k}\cdot\underline{x}\;,\;\lambda_2x_2\;,\;...\;,\;\lambda_nx_n\;\Big)\;.
\end{align}
An alternative unfolding of \cref{primaryunfold} can be written for $n=r+1$ as
\begin{align}
\v F=\Big(\;\psi(x_1)+\underline{k}\cdot\underline{x}\;,\;\lambda_2(x_2-\alpha_1)\;,\;...\;,\;\lambda_n(x_n-\alpha_{r})\;\Big)
\end{align}
where
\begin{align}
\psi(x_1)=x_1^{r+1}+\sum_{i=1}^{r}x_{i+1}x_1^{i-1}\;,
\end{align}
in which $\psi(x_1)$ is then the normal form of a Morin singularity of the mapping $$(x_1,x_2,...,x_n)\mapsto\bb{\psi(x_1),x_2,,...,x_n}\;.$$
We can generalize this to $n>r+1$ by just adding to this $F_{i}=x_i$ for $i>r+1$. Again the functions $\op B_r$ simplify to the same one dimensional derivatives \cref{mB2} as the primary form \cref{primary}. Morin singularities are a special case of singularities within Thom's much more general classification, namely those with Thom-Boardman symbols $\cc{i_1,i_2,...,i_r}=\cc{1,1,...,1}$. The Thom-Boardman symbols for mappings in $\mathbb R^m\to\mathbb R^n$ are a non-increasing sequence of integers $\cc{i_1,i_2,...}$, where each $i_j$ is the rank of a Jacobian matrix consisting of $\v F$ and of its minors of higher orders, taken via an iterative procedure as set out in \cite{arnold85,boardman67,thom56,thom55}. While this and other methods (see e.g. \cite{gaffney83,saji09}) may permit one to classify the type of singularity in a mapping at a given point, as we have noted, our aim here is more pragmatically to first find {\it where} any given bifurcation event may occur, before such standard methods can then be used to determine its precise classification.

\section{The catastrophes: examples}\label{sec:examples}

Below we use the conditions in \cref{def:sings}, which we will refer to as the `\BG {\it conditions}' for convenience, to find catastrophes up to codimension 6 in some planar vector fields. 
We include some simpler examples in an Appendix to illustrate the calculation and limitations of the \BG conditions, along with some novel geometries. 

The index notation used in previous sections is convenient for general formulae, but for easier reading in the following examples we use variables 
$\v x=(x,y,z,...)$, parameters 
$\alphab=(\alpha,\beta,...)$, and functions 
$\v F=(f,g,h,...)$. 
A term written $\op B_*$ denotes that $\op B$ is evaluated at a critical point $\v x_*$.

\subsection{Swallowtails ($r=3$): a cubic-quadratic form}\label{sec:r3}

An obvious form of swallowtail catastrophe would be
\begin{align*}
(f,g)&=\bb{y+x^4+\alpha x^2+\beta x+\gamma,\;y}\;,
\end{align*}
and one can imagine numerous generalisations where the relation to Arnold's principal families or Thom's one-dimensional catastrophes is obvious. 
Let us instead take a less trivial example, and show that it has a swallowtail catastrophe without performing any dimension reduction.


Consider the system
\begin{align}\label{swallowdouble}
(f,g)&=\bb{y+x^3+\alpha x+\beta,\;y^2+x+\gamma}\;,
\end{align}
which could have up to six critical points, which we label $(x_*,y_*)$. At these we evaluate the \BG conditions,
\begin{align}
\op B_1{}_*&=\abs{\sfrac{\partial(f,g)}{\partial{(x,y)}}}_*=\dabss{\begin{array}{cc}\alpha+3x^2&1\\1&2y\end{array}}_*=-1+2\alpha y_*+6x_*^2y_*\;,\nonumber\\
\op B_2{}_*&=\abs{\sfrac{\partial(\op B_1,g)}{\partial{(x,y)}}}_*=\dabss{\begin{array}{ccc}12xy&2\alpha+6x^2\\1&2y\end{array}}_*=24x_*y_*^2-6x_*^2-2\alpha\;,\nonumber\\
\op B_3{}_*&=\abs{\sfrac{\partial(\op B_2,g)}{\partial{(x,y)}}}_*=\dabss{\begin{array}{cccc}24y_*^2-12x_*&48x_*y_*\\1&2y\end{array}}_*=24y_*(2y_*^2-3x_*)\;,
\end{align}
which we can solve to find that these $\op B_i$ all vanish at 
\begin{align}\label{swal1}
(x,y)=\bb{\sfrac{1}{2^{3/5}3},\sfrac{1}{2^{4/5}}}\;,\qquad (\alpha,\beta,\gamma)=\bb{\sfrac{5}{2^{1/5}6},\sfrac{-35}{2^{4/5}27},\sfrac{-5}{2^{3/5}6}}\;.
\end{align}
This is a swallowtail catastrophe, and is full because the degeneracy conditions are non-zero, 
\begin{align}
\op G_{3,11}{}_*&=\abs{\sfrac{\partial(f,g,\op B_1,\op B_{2,1},\op B_{3,11})}{\partial{(x,y,\alpha,\beta,\gamma)}}}_*=720\;,\nonumber\\
\op G_{3,12}{}_*&=\abs{\sfrac{\partial(f,g,\op B_1,\op B_{2,1},\op B_{3,12})}{\partial{(x,y,\alpha,\beta,\gamma)}}}_*=-2^{4/5}360\;,\nonumber\\
\op G_{3,21}{}_*&=\abs{\sfrac{\partial(f,g,\op B_1,\op B_{2,2},\op B_{3,21})}{\partial{(x,y,\alpha,\beta,\gamma)}}}_*=2^{3/5}360\;,\nonumber\\
\op G_{3,22}{}_*&=\abs{\sfrac{\partial(f,g,\op B_1,\op B_{2,2},\op B_{3,22})}{\partial{(x,y,\alpha,\beta,\gamma)}}}_*=-2^{2/5}360\;.
\end{align}
Emanating from the swallowtail point in $(\alpha,\beta,\gamma)$ space, there are curves of cusps where $\op B_1{}_*=\op B_2{}_*=0$, and a surface of folds where $\op B_1{}_*=0$, each with suitable non-degeneracy conditions, only violated at the higher codimension catastrophes. 

\begin{figure}[h!]\centering
\includegraphics[width=0.49\textwidth]{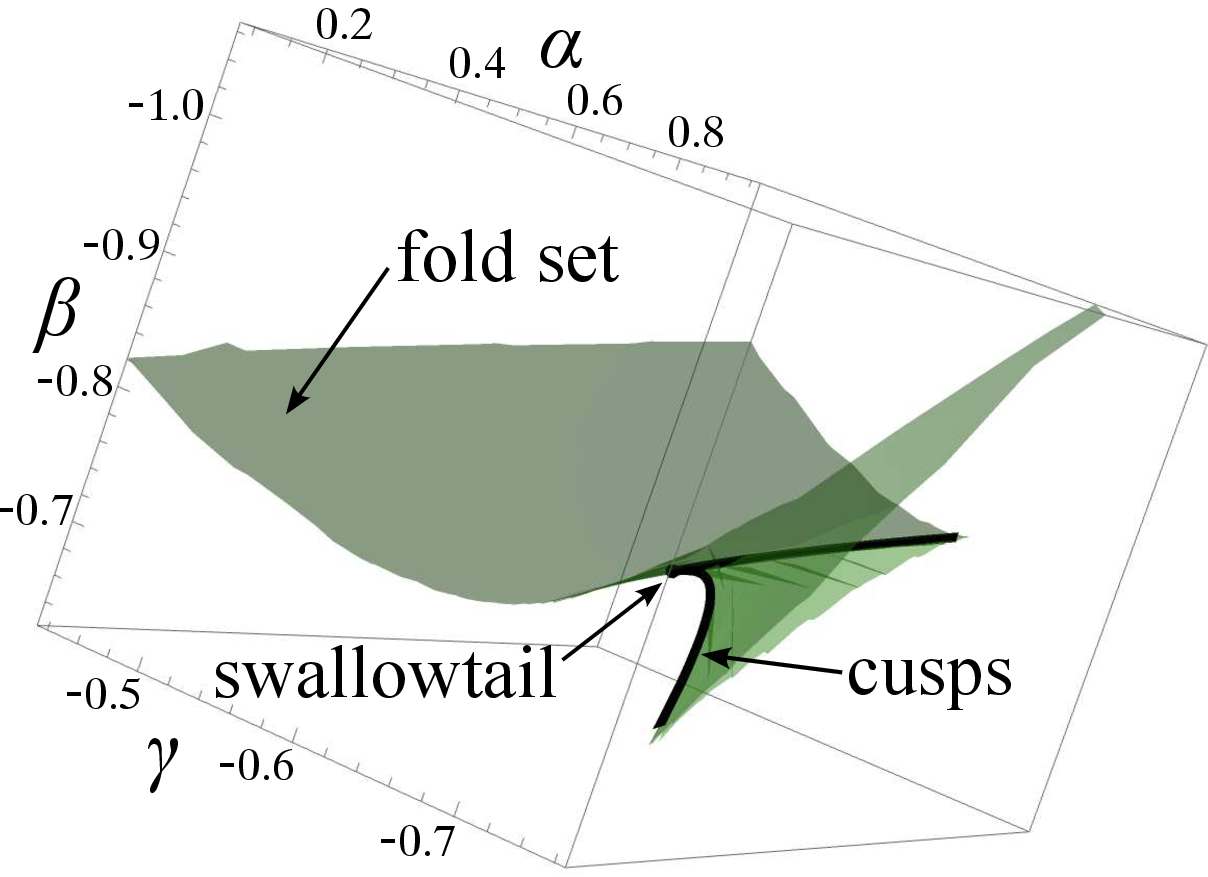}
\includegraphics[width=0.49\textwidth]{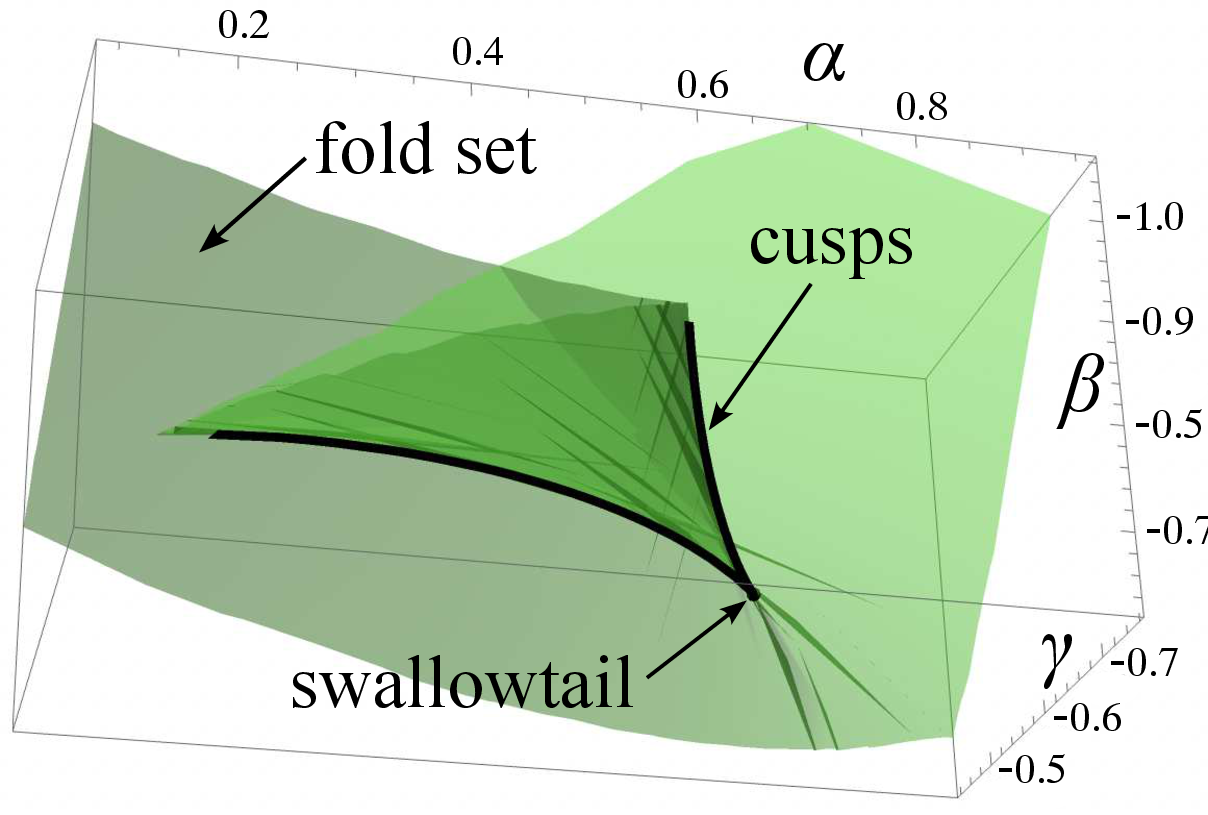}
\vspace{-0.3cm}\caption{\small\sf The fold, cusp, and swallowtail geometry of the system \cref{swallowdouble} shown from two directions. In $(\alpha,\beta,\gamma)$ parameter space the folds lie on the surface which shows the characteristic shape of a swallowtail, with two curves of cusps, meeting at the swallowtail point. 
}\label{fig:swal}\end{figure}

It is not easy to express the fold and cusp sets explicitly, but we can find them parametrically using the \BG conditions, resulting in the illustration in \cref{fig:swal}. 
The fold set can be parameterised by setting $x=s$, $y=t$, as parameters, and seeking a solution of $f=g=\op B_1=0$ in terms of these, giving
\begin{align}
\alpha=(1-6s^2t)/2t\;,\quad \beta=2s^3-t-\sfrac{s}{2t}\;,\quad \gamma=-t^2-s\;.
\end{align}
Solving also $\op B_2=0$ we find the cusps appear at $s=1/24t^3$, hence they are given parametrically by
\begin{align}
\alpha=\sfrac{1}{2t}-\sfrac1{192t^6}\;,\quad\beta=\sfrac1{6912t^9}-\sfrac1{48t^4}-t\;,\quad \gamma=-\sfrac1{24t^3}-t^2\;.
\end{align}
Finally solving $\op B_3=0$ we find the swallowtail then lies at $t=2^{-4/5}$ (which gives coordinate and parameter values agreeing with \cref{swal1}). These are plotted in \cref{fig:swal}, revealing the familiar geometry of the swallowtail catastrophe, where a fold surface kinks along a pair of cusp curves that intersect at the swallowtail.

\subsection{Butterflies ($r=4$)}\label{sec:r4}

Similarly to the previous section, one can imagine numerous obvious ways of embedding the known normal forms of butterfly catastrophes in higher dimensions, for example 
\begin{align*}
(f,g)&=\bb{y+x^5+\alpha x^3+\beta x^2+\gamma x,\;y-\delta}\;.
\end{align*}
However, as in the last section, let us look at a more novel form, followed by the application that inspired it.


Consider the vector field
\begin{align}\label{butterflycubic}
(f,g)&=\bb{y+x^3+\alpha x+\beta,\;x+ y^3+\gamma y+\delta}\;,
\end{align}
which could, based on its order, have potentially up to 9 critical points $(x_*,y_*)$. The determinants $\op B_j$ evaluate as
\begin{align}
\op B_1{}_*&=\abs{\sfrac{\partial(f,g)}{\partial{(x,y)}}}_*
=(\alpha+3x_*^2)(\gamma+3y_*^2)-1\;,\nonumber\\
\op B_2{}_*&=\abs{\sfrac{\partial(\op B_1,g)}{\partial{(x,y)}}}_*=6x_*(\gamma+3y_*^2)^2-6y_*(\alpha+3x_*^2)\;,\nonumber\\
\op B_3{}_*&=\abs{\sfrac{\partial(\op B_2,g)}{\partial{(x,y)}}}_*=6(\alpha+3x_*^2)+6(\gamma+3y_*^2)\bb{(\gamma+3y_*^2)^2-18x_*y_*}\;,\nonumber\\
\op B_4{}_*&=\abs{\sfrac{\partial(\op B_3,g)}{\partial{(x,y)}}}_*=72\bb{15x_*y_*^2-3\gamma^2y_*-27y_*^5+2\gamma(x_*-9y_*^3)}\;.
\end{align}
These all vanish, indicating butterfly catastrophes, at 
\begin{align}
x=y=\pm\sfrac13\;,\qquad \alpha=\gamma=\sfrac23\;,\quad \beta=\delta=\mp\sfrac{16}{27}\;,
\end{align}
with non-degeneracy conditions that all evaluate as 
$\op G_{4,i_1i_2i_3}=\pm103680$ for any $i_1,i_2,i_3\in\cc{1,2}$, hence they are full. 
Emanating from the butterfly point in $(\alpha,\beta,\gamma,\delta)$ space, there are curves of swallowtails where $\op B_1{}_*=\op B_2{}_*=\op B_3{}_*=0$, surfaces of cusps where $\op B_1{}_*=\op B_2{}_*=0$, and a volume of folds where $\op B_1{}_*=0$, each with suitable non-degeneracy conditions that are only violated at the higher codimension catastrophes. 


As in the previous section, we can find a parameterisation of these geometries using the \BG conditions. To find the folds, solving $f=g=\op B_1=0$ for $(\alpha,\beta,\delta)$, and taking as parameters $x=s$, $y=t$, and $\gamma$, we find 
\begin{align}
\alpha=\sfrac{1}{\gamma+3t^2}-3s^2\;,\quad
\beta=2s^3-t-\sfrac{s}{\gamma+3t^2}\;,\quad
\delta=-s-\gamma t-t^3\;.
\end{align}
To find the cusps, solving also $\op B_2=0$ for $\gamma$ in terms of the same parameters we find $\gamma=\sfrac{t^{1/3}}{s^{1/3}}-3t^2$, giving
\begin{align}
\alpha=a(s,t)\;,\quad
\beta=b(s,t)\;,\quad
\gamma=a(t,s)\;,\quad
\delta=b(t,s)\;,
\end{align}
where
\begin{align}
a(s,t)=\sfrac{s^{1/3}}{t^{1/3}}-3s^2\;,\quad
b(s,t)=2s^3-t-\sfrac{s^{4/3}}{t^{1/3}}\;.
\end{align}
The swallowtail is slightly more difficult to solve for. If we let $(s,t)=\big((uv)^{1/2},(v/u)^{1/2}\big)$, with inverse $(u,v)=(s/t,st)$, then $\op B_3=0$ has the solution $v=(u^{2/3}+u^{-2/3})/18$, hence we have swallowtail curves parameterised by $u$ as
\begin{align}
\alpha&=a\bb{(uv)^{1/2},(v/u)^{1/2}}\;,\quad
\beta=b\bb{(uv)^{1/2},(v/u)^{1/2}}\;,\nonumber\\
\gamma&=a\bb{(v/u)^{1/2},(uv)^{1/2}}\;,\quad
\delta=b\bb{(v/u)^{1/2},(uv)^{1/2}}\;,\\ 
v&=(u^{2/3}+u^{-2/3})/18\;.\nonumber
\end{align}
Lastly the butterfly is found by solving $\op B_4=0$, which happens simply when $u=1$, giving
\begin{align}
\alpha=\gamma=\sfrac23\;,\quad
\beta=\delta=\pm\sfrac{16}{27}\;.
\end{align}
We delay illustrations of this case to the more general scenario in the next section.

\subsection{A motivating PDE example: stars and butterflies}\label{sec:pde}

The search for the \BG conditions in this paper was motivated by an application to a PDE system for modeling polarity in cell structures \cite{j21fahad} (which includes bifurcation diagrams found by numerical continuation of the \BG conditions). Schematically this took the form
\begin{align}
\sfrac{\partial\;}{\partial t}(A,B)&=\sfrac{\partial^2\;}{\partial u^2}(A,B)+(f,g)\;,
\end{align}
where
\begin{align}\label{cubicpde}
(f,g)&=\bb{\rho y+x^3+\alpha x+kxy+\beta,\;\sigma x+ y^3+\gamma y+kxy+\delta}\;.
\end{align}
The problem of interest in the application is to study both local and global spatiotemporal patterns, specifically heteroclinic connections between homogeneous steady states where $(f,g)=(0,0)$, which exist in regions bounded by catastrophes of the vector field $(f,g)$. 

The system \cref{cubicpde} is just a slight generalization of the double-cubic butterfly in \cref{butterflycubic}, but its analysis is made rather more difficult by the non-gradient terms $kxy$ and $\rho\neq\sigma$. Nevertheless, like the system in \cref{butterflycubic}, this is invariant under the symmetry $$(x,y,\alpha,\beta,\gamma,\delta,\rho,\sigma)\mapsto(y,x,\gamma,\delta,\alpha,\beta,\sigma,\rho)\;,$$ and this can be exploited to help find the critical points and catastrophes. 

One can use the \BG conditions to find the full set of catastrophes numerically. We can also solve for them analytically if we restrict ourselves slightly, for example to looking for catastrophes that lie on the part of the invariant surface $x=y$, $\rho=\sigma$. Then solving $f=g=\op B_1=0$, letting $x=s$, $y=t$, be parameters, we find the fold set is given by
\begin{align}
\alpha&=\sfrac{s\cc{\rho^2+3\rho s(k+s)+\delta(k+3s)+s^2(k^2-2ks-6s^2)}}{2s^3-\rho s-\delta}\;,\nonumber\\
\beta&=\sfrac{\delta s(\rho-2s^2)-s^3(k+2s)(2\rho+ks-2s^2)}{2s^3-\rho s-\delta}\;,\nonumber\\ 
\gamma&=-\rho-s(k+s)-\sfrac\delta s\;,
\end{align}
and this is full since $\op G_1$ (which we will not write here for brevity) does not vanish at typical $s,\rho,\delta,k$. 

Solving also $\op B_2=0$ we find cusps at
\begin{align}
\alpha=\gamma=\rho-3s^2\;,\quad
\beta=\delta=s\cc{s(2s-k)-2\rho}\;.
\end{align}
Here we find $\op G_{2,i}=(-1)^i72(\rho+k s)^2(\rho+4ks-9s^2)s$ for $i=1,2$, so the cusps are full at typical values of $s,\rho,k$. 

Solving $\op B_3=0$ we find swallowtails occur where $\rho=s(9s-4k)$, but having made the assumptions $x=y$ and $\rho=\sigma$ the swallowtails turn out to be degenerate, with $\op G_{3,i_1i_2}=0$ for all $i_1,i_2\in\cc{1,2}$. (Full swallowtails may occur  away from this symmetry set where $x\neq y$ and/or $\rho\neq\sigma$). 

Evaluating $\op B_4=0$ we find that we have actually located a {\it butterfly} at $\rho=s(9s-4k)$, hence these are given parametrically by 
\begin{align}
\alpha=\gamma=2s(3s-2k)\;,\quad
\beta=\delta=s^2(16s-7k)\;,\quad
\rho=s(9s-4k)\;.
\end{align}
The non-degeneracy conditions all evaluate as $\op G_{4,i_1i_2i_3}=\pm 2^73^85s^6(k-3s)^7(k - 9s) (2 k - 9s) (k - 6s)$ for any $i_1,i_2,i_3\in\cc{1,2}$, so the butterflies are full at typical values of $s,k$. 

Since $\op G_{4,I}$ can vanish at certain points, this suggests there are higher codimension catastrophes here too. If we now solve $\op B_5=0$ to look for a {\it wigwam} catastrophe, we find solutions $\rho=-ks$ and $\rho=s(9s-4k)$. 
Again, having made the assumptions $x=y$ and $\rho=\sigma$, both of these are degenerate as $\op G_{5,i_1i_2i_3i_4}=0$ for all $i_1,i_2,i_3,i_4\in\cc{1,2}$. 

As before we have actually found a higher codimension point here, a {\it star} catastrophe where $\op B_6=0$, therefore lying at 
\begin{align}
\alpha=\gamma=-\hf k^2\;,\quad\beta=\delta=\sfrac{13}{108}k^3\;,\quad\rho=\sigma=-\sfrac5{12}k^2\;.
\end{align}
This is full for $k\neq0$, with $\op G_{6,i_1i_2i_3i_4i_5}=\pm\sfrac{3^85^27}{2^{15}}k^{29}$ for any $i_1,i_2,i_3,i_4,i_5\in\cc{1,2}$.

\begin{figure}[h!]\centering\includegraphics[width=0.8\textwidth]{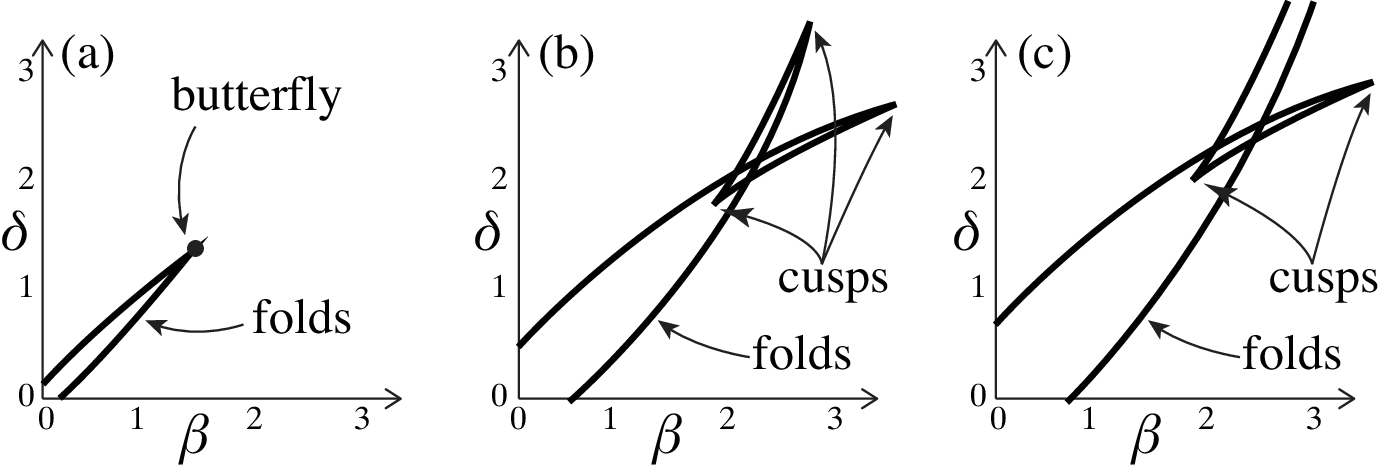}
\vspace{-0.3cm}\caption{\small\sf A bifurcation diagram for the system \cref{cubicpde} in $(\beta,\delta)$ space with $k=1$, $r=s=7/3$, and: (a) $\alpha=\gamma=2$, (b) $\alpha=\gamma=1.6$, (c) $\alpha=1.2$, $\gamma=1.6$.
}\label{fig:fbut}\end{figure}

A depiction of the fold surfaces, cusps, and butterfly, plotted in the $(\beta,\delta)$ plane, is shown in \cref{fig:fbut} for the given parameters. A more detailed study of this system along with its applications can be found in \cite{j21fahad}. 

%


\section{Closing Remarks}\label{sec:conc}

The \BG conditions can be thought of as a geometric test. The functions $\op B_r$ characterise the order of contact between the hypersurfaces $F_i=0$ as they intersect to form a critical point $\v F=0$. They are therefore a means to {\it locate} bifurcation points, but as they contain no information about directionality of $\v F$, they do not define bifurcation equivalence classes. Instead they define an underlying catastrophe, at which a bifurcation of the vector field occurs which may unfold via a number of non-equivalent routes. For example, if the \BG conditions reveal a swallowtail involving four equilibria in a vector field, then the bifurcation there could be a simple swallowtail,  involving a bifurcation of four equilibria with no other local topological degeneracies, or it could involve homoclinic or heteroclinic connections and limit cycles, which require further local stability analysis beyond the underlying catastrophe. Once the underlying catastrophe is located, its stability and topology can be studied using standard local methods to fully define its bifurcation class. 

The conditions of being {\it full} are necessary to guarantee that the catastrophe conditions have unique solutions. If the $\op G$ conditions are violated, then the problem $\v F=\op B_1=...=\op B_r=0$ may be unsolvable for some choices of the $\op B_{i,j}$ (recall in \cref{def:sings} we are able to select $\op B_{i,I(i-1)}=\op B_{i,1...1}$ without loss of generality), and then closer study is required to find the nature of the degeneracy. This is one reason why the relation between these conditions, and the classifications of singularity theory, are not yet fully understood; in seeking more practicality we have had to give up some generality. 

Hence it remains to develop a theory relating the \BG conditions to the known classifications of singularity theory, to understand how they translate into certain germs, Thom-Boardman symbols, and versality conditions, and ultimately determine whether the \BG conditions at codimension $r$ are sufficient to imply that a point 
breaks into $r+1$ critical points under perturbation. The geometrical insight used here is rather different to the more powerful but less directly calculable classifications of singularity theory, namely that rather than characterize a singularity via its equivalence to a germ and the ranks of certain ideals, we interpret a critical point as an intersection of certain hypersurfaces and translate this into a readily calculable set of \BG conditions. It is not immediately obvious how to relate these two approaches, but is no doubt possible. 

In certain circumstances, a vanishing of some of the $\op G$ conditions may constitute a trivial violation of the property of being `full', which may be resolved by eliminating one or more redundant dimensions (see \cref{sec:cuspsimple}). It may be possible to derive a test for this trivial form of degeneracy.

Although the catastrophe conditions are derived from the geometry of the nullclines $F_i=0$, which are coordinate dependent, the functions $\op B_i$ are determinants expressing linear independency of the components of $\v F$, and are therefore coordinate {\it in}dependent. 



%
%

\bigskip\noindent{\bf Acknowledgements. }
My thanks to Fahad Al Saadi and Alan Champneys, whose request for a formula to detect a suspected swallowtail catastrophe in a biological reaction diffusion model inspired this work --- the catastrophe turned out to be a butterfly and is shown in a schematic form in \cref{sec:pde}. 

\bigskip\bigskip

\noindent{\Large\bf Appendix}

\appendix

\section{The \BG functions for low codimension}\label{sec:BGexpanded}

For ease of reference, the functions defined in \cref{Br} and \cref{Gri} are as follows for the first few codimensions $r$. 

For $r=1$ the string $I(r)$ is undefined, and we have simply
\begin{subequations}
\begin{align}
		\op B_1&=\abs{\frac{\partial(F_1,...,F_{n})}{\partial(x_1,...,x_n)}}\;,\\\
		\op G_{1}&=\abs{\frac{\partial(F_1,...,F_{n},\op B_{1})}{\partial(x_1,...,x_n,\alpha_1)}}\;.\qquad\qquad\qquad
\end{align}
\end{subequations}

For $r=2$ we have $I(1)=i_1$, with $\op B_1$ as defined above, and
\begin{subequations}
\begin{align}
		\op B_2&=\abs{\frac{\partial(\op B_{1},F_2,...,F_{n})}{\partial(x_1,...,x_n)}}\;,\\
		\op G_{2,i_1}&=\abs{\frac{\partial(F_1,...,F_{n},\op B_{1},\op B_{2,i_1})}{\partial(x_1,...,x_n,\alpha_1,\alpha_{2})}}\;,\\
		\op B_{2,i_1}&=\abs{\frac{\partial(F_1,...,F_{i_1-1},\op B_{1},F_{i_1+1},...,F_n)}{\partial(x_1,...,x_n)}}\;,
\end{align}
\end{subequations}
		for $i_1\in\cc{1,...,n}$. 

For $r=3$ we have $I(2)=i_1i_2$, with $\op B_1,\op B_2,$ and $\op B_{2,i_1}$ as defined above, and
\begin{subequations}
\begin{align}
		\op B_3&=\abs{\frac{\partial(\op B_{2},F_2,...,F_{n})}{\partial(x_1,...,x_n)}}\;,\\
		\op G_{3,i_1i_2}&=\abs{\frac{\partial(F_1,...,F_{n},\op B_{1},\op B_{2,i_1},\op B_{3,i_1i_2})}{\partial(x_1,...,x_n,\alpha_1,\alpha_2,\alpha_{3})}}\;,\\
		\op B_{3,i_1i_2}&=\abs{\frac{\partial(F_1,...,F_{i_2-1},\op B_{2,i_1},F_{i_2+1},...F_n)}{\partial(x_1,...,x_n)}}\;,
\end{align}
\end{subequations}
		for $i_1\in\cc{1,...,n}$ and $i_2\in\cc{1,...,n}$. 

So it continues. For $r=4$ we have the functions $\op B_1$ to $\op B_4$, along with $\op G_{4,i_1i_2i_3}$ in terms of $\op B_{4,i_1i_2i_3}$, $\op B_{3,i_1i_2}$, and $\op B_{2,i_1}$. As stated in \cref{sec:BG}, it remains for general $r$ to establish whether all of the $n^{r-1}$ functions $\op G_{r,i_1...i_{r-1}}$ are independent, and therefore whether they are all necessary to evaluate, but their calculation is at least straightforward.

\section{Further examples: folds and cusps}\label{sec:further}

Although folds and cusps are relatively easy to study using conventional theory, we include a few examples here to further illustrate the calculation of the \BG conditions in \cref{def:sings}. 
For a fold in particular, the condition $\op B_1=0$ is just the well known condition that the system's Jacobian has less than full rank at a singularity, so our interest lies mainly in interpreting the non-degeneracy conditions. 

In some cases we apply the conditions to a dynamical system $(\dot x,\dot y,...)=(f,g,...)$ to highlight the significance of ignoring the directionality of a vector field.



\subsection{Simple fold}\label{sec:foldsimple}

\begin{figure}[h!]\centering\includegraphics[width=0.19\textwidth]{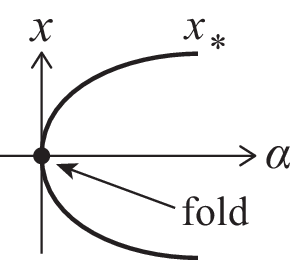}
\vspace{-0.3cm}\caption{\small\sf A simple fold where two branches of critical points coalesce. 
}\label{fig:fold}\end{figure}

Consider the vector field 
\begin{align}
(f,g)&=(x^2-y,\;y-\alpha)\;.
\end{align}
The critical points lie at $(x_*,y_*)=(\pm\sqrt{\alpha},\alpha)$, at which the \BG conditions evaluate as
\begin{subequations}
\begin{align}
\op B_1{}_*&=\abs{\sfrac{\partial(f,g)}{\partial{(x,y)}}}_*=\dabss{\begin{array}{cc}2x&-1\\0&1\end{array}}_*=\pm2\sqrt\alpha\;,\\
\op G_1{}_*&=\abs{\sfrac{\partial(f,g,\op B_1)}{\partial{(x,y,\alpha)}}}_*=\dabss{\begin{array}{ccc}2x&-1&0\\0&1&-1\\2&0&0\end{array}}_*=2\;.
\end{align}
\end{subequations}
We see that $\op B_1{}_*$ vanishes at $\alpha=0$, and this is a full fold since $\op G_1{}_*\neq0$. The familiar bifurcation diagram is shown in \cref{fig:fold}. In a dynamical system $(\dot x,\dot y)=(f,g)$ this is a degenerate equilibrium that perturbs into a saddle and node.

If we swap $(f,g)\mapsto(g,f)$ or $(f,g)\mapsto(f,-g)$, the \BG conditions are unchanged, and this remains a full fold. These vector fields have very different directional properties, though, and are only {\it topologically} equivalent up to a time reversal. This is because the \BG conditions detect only the underlying bifurcation and neglect directionality of the vector field. This is particularly well highlighted if we take a similar vector field
\begin{align}
(f,g)&=(y,\;x^2-\alpha)\;.
\end{align}
Here the fold that occurs is between a saddle and a center, as $(f,g)$ is divergence-free, and therefore is not topologically equivalent to the cases above. Nevertheless, by classifying what happens at $x=y=\alpha=0$ as a {\it fold} catastrophe underlying the topological and stability features of the vector field, we are able to locate the bifurcation point using the \BG conditions.

\subsection{Generalised codimension 3 fold of {\it Dumortier, Roussarie, \& Sotomayor}}\label{sec:dumortier}

There are various non-trivial generalisations of the fold bifurcation in the singularity theory literature, for example the codimension 2 `fold-Hopf' or `zero-Hopf' bifurcation. Let us take an example from \cite{drs87}, a codimension 3 fold-Hopf bifurcation with normal form
\begin{align}\label{drs}
(f,g)&=\bb{y,\;x^2-\alpha+y(k_0+k_1x+x^3)}\;.
\end{align}
This has critical points $(x_*,y_*)=(\pm\sqrt\alpha,0)$. Taking $\alpha$ as the bifurcation parameter, we have
\begin{subequations}
\begin{align}
\op B_1{}_*&=\abs{\sfrac{\partial(f,g)}{\partial{(x,y)}}}_*=\dabss{\begin{array}{cc}0&1\\2x+ y(3x^2+k_1)&k_0+k_1x+x^3\end{array}}_*=\pm2\sqrt\alpha\;,\\
\op G_1{}_*&=\abs{\sfrac{\partial(f,g,\op B_1)}{\partial{(x,y,\alpha)}}}_*=\dabss{\begin{array}{ccc}0&1&0\\2x+\alpha y(3x^2+k_1)&k_0+k_1x+x^3&-1\\-2-6xy&-k_1-3x^2&0\end{array}}_*=2\;.
\end{align}
\end{subequations}
This therefore satisfies the conditions of a full fold at $\alpha=0$. The higher order bifurcation studied by {\it Dumortier et al.} in \cite{drs87} is generated by the $y(k_0+k_1x+x^3)$ term in \cref{drs}, and is associated with a loss of stability that occurs where the trace of the Jacobian, $k_0\pm k_1\alpha^{1/2}\pm\alpha^{3/2}$, vanishes, around which can occur, in the language of a dynamical system $(\dot x,\dot y)=(f,g)$, Hopf bifurcations, homoclinic connections, and a saddle whose eigenvalues have equal magnitude. 
\begin{figure}[h!]\centering\includegraphics[width=0.68\textwidth]{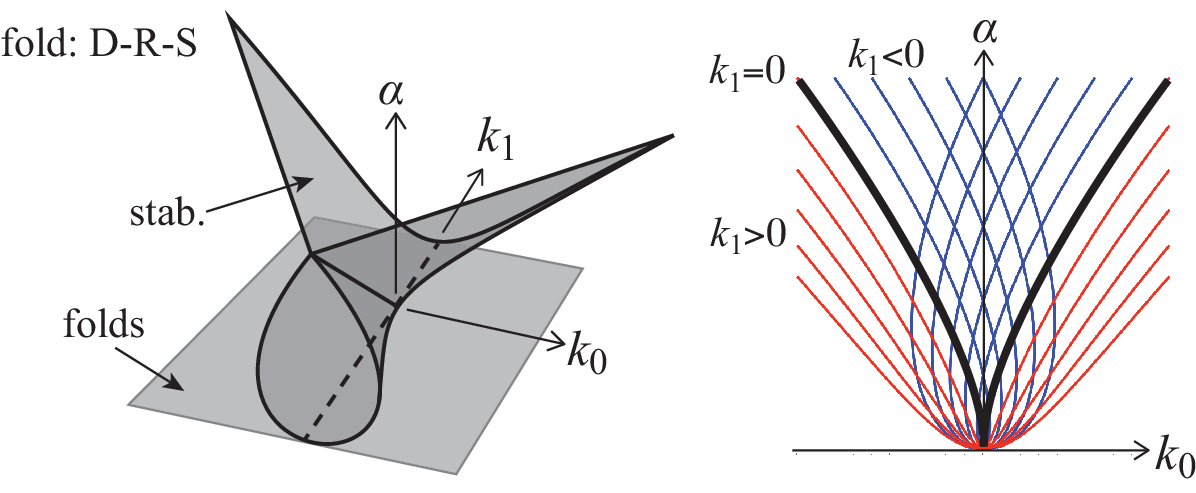}
\vspace{-0.3cm}\caption{\small\sf Bifurcation sets of the codimension 3 fold, showing the fold set where $\op B_1{}_*=0$ and the set `stab' where $\tr\bb{\sfrac{d\v F}{d\v x}}{}_*=0$. Right: sections of the `stab' surface for $k_1>0$ (red) and $k_1<0$ (blue), with a cusp at $k_1=0$.   
}\label{fig:dumortier}\end{figure}

These sets are illustrated in parameter space in \cref{fig:dumortier}. Two branches of the `stab' surface intersect along the line $k_0=k_1+\alpha=0$. A codimension 3 catastrophe occurs at the origin. See \cite{drs87} for in-depth analysis of this singularity, where it is referred to as a `cusp' bifurcation due to the geometry of the stability curve, while here we classify it as a fold in keeping with it being a `fold' of the critical points $x_*=\pm\sqrt\alpha$. The \BG conditions detect the underlying fold, irrespective of the stability-related bifurcation structure on top.

\subsection{Bogdanov-Takens bifurcation}\label{sec:bogdanovtakens}

\begin{figure}[h!]\centering\includegraphics[width=0.27\textwidth]{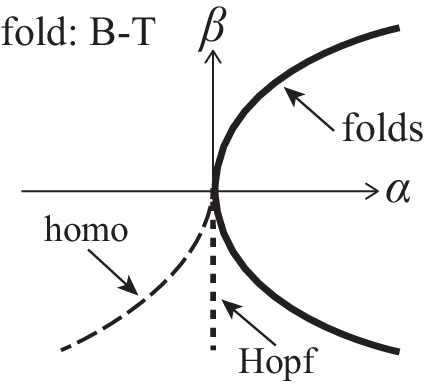}
\vspace{-0.3cm}\caption{\small\sf Bifurcation diagram of the Bogdanov-Takens bifurcation in $(\alpha,\beta)$ space, showing the curve of folds, Andronov-Hopf bifurcations, and homoclinic bifurcations. 
}\label{fig:bogdanov}\end{figure}

The vector field 
\begin{align}
(f,g)&=(y,\;\alpha+\beta x+x^2-xy)\;
\end{align}
has critical points $(x_*,y_*)=(\hf(-\beta\pm\sqrt{\beta^2-4\alpha}),0)$. Let us consider taking either $\alpha$ or $\beta$ as a bifurcation parameter for a fold, and evaluate
\begin{subequations}
\begin{align}
\op B_1{}_*&=\abs{\sfrac{\partial(f,g)}{\partial{(x,y)}}}_*=\dabss{\begin{array}{cc}0&1\\\beta+2x-y&-x\end{array}}_*
=\mp\sqrt{\beta^2-4\alpha}\;,\\
\op G_1^{\alpha}{}_*&=\abs{\sfrac{\partial(f,g,\op B_1)}{\partial{(x,y,\alpha)}}}_*=\dabss{\begin{array}{ccc}0&1&0\\\beta+2x-y&-x&1\\-2&1&0\end{array}}_*=-2\;,\nonumber\\
\op G_1^{\beta}{}_*&=\abs{\sfrac{\partial(f,g,\op B_1)}{\partial{(x,y,\beta)}}}_*=\dabss{\begin{array}{ccc}0&1&0\\\beta+2x-y&-x&x\\-2&1&-1\end{array}}_*=\beta\;.
\end{align}
\end{subequations}
A fold occurs along the curve $\alpha=\beta^2/4$ in parameter space. The function $\op G_1^{\alpha}$ indicates that the fold is full if we vary $\alpha$. If we vary $\beta$, however, then $\op G_1^{\beta}$ indicates that the fold is full except at $\beta=\alpha=0$, and as we see in \cref{fig:bogdanov}, this occurs because varying $\beta$ will give a path through parameter space that tangentially `bounces off' the fold curve if $\alpha=0$. 

As before, the \BG conditions reveal the underlying fold catastrophe. In this case a stability change also creates, taking a dynamical system $(\dot x,\dot y)=(f,g)$, an Andronov-Hopf bifurcation along the half-line $\alpha=0$, $\beta<0$, producing a limit cycle that disappears in a homoclinic bifurcation along a curve $\alpha\approx-\sfrac6{25}\beta^2$, $\beta<0$; see e.g. \cite{bogdanov81,k98,takens74} for the local analysis. The point $\alpha=\beta=0$ is of course the Bogdanov-Takens bifurcation.

\subsection{Fold set as a Whitney umbrella}\label{sec:foldwhitney}

\begin{figure}[h!]\centering\includegraphics[width=0.7\textwidth]{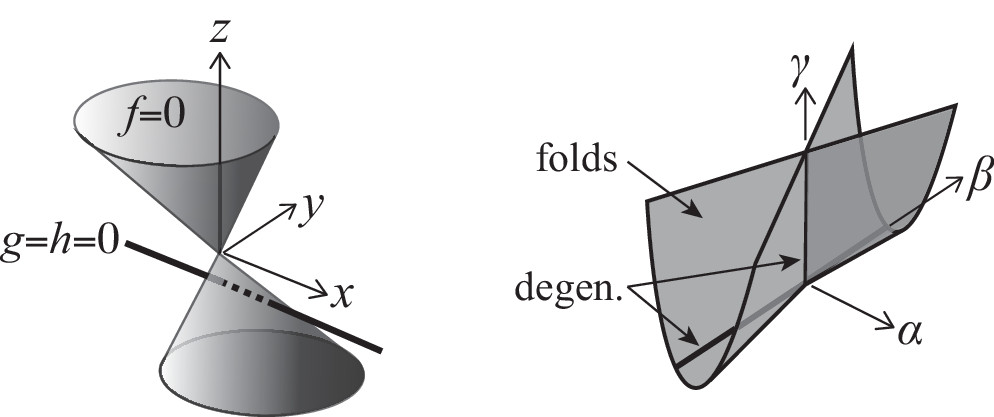}
\vspace{-0.3cm}\caption{\small\sf The double conic set $f=0$ and line $g=h=0$ (left), and the fold set forming a Whitney umbrella (right). 
}\label{fig:whitney}\end{figure}

The vector field
\begin{align}
(f,g,h)=(x^2+2zx+cy^2,\;y-\beta,\;z-\alpha)
\end{align}
has critical points at $(x_*,y_*,z_*)=(-\alpha\pm\sqrt{\alpha^2-\beta^2\gamma},\beta,\alpha)$. This is a contrived example but shows some novel geometry as seen in \cref{fig:whitney}. 
The hypersurface $f=0$ is a double conic surface, forming equilibria where it is pierced by the straight line given by $y-\beta=z-\alpha=0$, and as either of these move we will observe folds. We have 3 parameters and so could investigate bifurcations with respect to either of them, and we would obtain similar results, so let us take $\alpha$ without loss of generality. 
The \BG conditions give 
\begin{subequations}
\begin{align}
\op B_1{}_*&=\abs{\sfrac{\partial(f,g,h)}{\partial{(x,y,z)}}}_*=\dabss{\begin{array}{ccc}2x+2z&2\gamma y&2x\\0&1&0\\0&0&1\end{array}}_*=\pm2\sqrt{\alpha^2-\beta^2\gamma}\;,\\
\op G_1{}_*&=\abs{\sfrac{\partial(f,g,h,\op B_1)}{\partial{(x,y,z,\alpha)}}}_*=\dabss{\begin{array}{cccc}2x+2z&2\gamma y&2x&0\\0&1&0&0\\0&0&1&-1\\2&0&2&0\end{array}}_*=4\alpha\;.
\end{align}
\end{subequations}
There is a fold where $\op B_1{}_*$ vanishes, along $\alpha^2-\beta^2\gamma=0$, forming a Whitney surface in $(\alpha,\beta,\gamma)$ space, and this is non-degenerate provided $\alpha\neq0$. The zero sets and the fold surface are shown in \cref{fig:whitney}. It is clear why the system is degenerate when $\alpha=0$, as the line $y-\beta=z-\alpha=0$ can then only intersect the double-conic $f=0$ at its apex.


\subsection{Simple cusp, and the property of being `full'}\label{sec:cuspsimple}

\begin{figure}[h!]\centering\includegraphics[width=0.23\textwidth]{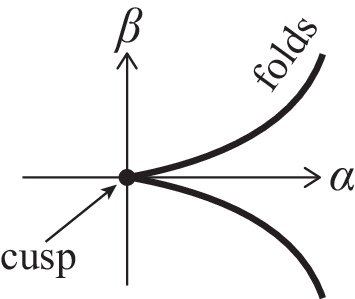}
\vspace{-0.3cm}\caption{\small\sf Bifurcation diagram of the simple cusp in $(\alpha,\beta)$ parameter space. }\label{fig:cusps}\end{figure}

Consider
\begin{align}
(f,g,h)=(y+x^3+\alpha x+\beta,\;y+kz,z)\;, 
\end{align}
which has critical points $(x_*,y_*,z_*)=(x_*,0,0)$, where $x_*$ are solutions of a cubic polynomial with discriminant $(\alpha/3)^3+(\beta/2)^2$.  The \BG conditions evaluate as
\begin{subequations}
\begin{align}
\op B_1{}_*&=\abs{\sfrac{\partial(f,g,h)}{\partial{(x,y,z)}}}_*=\dabss{\begin{array}{ccc}\alpha+3x^2&1&0\\0&1&k\\0&0&1\end{array}}_*=\alpha+3x_*^2\;,\nonumber\\
\op B_2{}_*&=\abs{\sfrac{\partial(\op B_1,g,h)}{\partial{(x,y,z)}}}_*=\dabss{\begin{array}{ccc}6x&0&0\\0&1&k\\0&0&1\end{array}}_*=6x_*\;,
%
%
\end{align}
with
\begin{align}
\op G_{2,1}{}_*&=\abs{\sfrac{\partial(f,g,h,\op B_1,\op B_{2,1})}{\partial{(x,y,z,\alpha,\beta)}}}_*=\dabss{\begin{array}{ccccc}\alpha+3x^2&1&0&x&1\\0&1&k&0&0\\0&0&1&0&0\\6x&0&0&1&0\\6&0&0&0&0\end{array}}_*=6\;,\nonumber\\
\op G_{2,2}{}_*&=\abs{\sfrac{\partial(f,g,h,\op B_1,\op B_{2,2})}{\partial{(x,y,z,\alpha,\beta)}}}_*=\dabss{\begin{array}{ccccc}\alpha+3x^2&1&0&x&1\\0&1&k&0&0\\0&0&1&0&0\\6x&0&0&1&0\\-6&0&0&0&0\end{array}}_*=-6\;,\nonumber\\
\op G_{2,3}{}_*&=\abs{\sfrac{\partial(f,g,h,\op B_1,\op B_{2,3})}{\partial{(x,y,z,\alpha,\beta)}}}_*=\dabss{\begin{array}{ccccc}\alpha+3x^2&1&0&x&1\\0&1&k&0&0\\0&0&1&0&0\\6x&0&0&1&0\\6k&0&0&0&0\end{array}}_*=-6k\;.
\end{align}
\end{subequations}
We omit the intermediate calculations of the determinants $\op B_{2,i}$. The familiar bifurcation diagram is shown in \cref{fig:cusps}.

As one would expect, the first condition $\op B_1{}_*$ vanishes to create folds on the curve $$(\alpha/3)^3+(\beta/2)^2=0$$ in parameter space, which are non-degenerate with $\op G_1^{\alpha}{}_*=\alpha-2x_*^2$ and $\op G_1^{\beta}{}_*=-6x_*$ (i.e. taken with respect to either $\alpha$ or $\beta$), both non-zero provided $\alpha,\beta\neq0$. 

The second bifurcation condition $\op B_2{}_*$ vanishes at $\alpha=\beta=0$, for which $x_*=0$ (at which we see the non-degeneracy conditions for the fold would also be violated). Note that all $\op G_{2,i}{}_*$ are non-vanishing, so the cusp is non-degenerate. 

Note, however, that non-degeneracy requires $k\neq0$. This is an important illustration of the property of being `full' in \cref{def:sings}, and stems from the gradient vectors, which evaluate at the cusp to $\sfrac{d\;}{d\v x}f=(0,1,0)$, $\sfrac{d\;}{d\v x}g=(0,1,k)$, $\sfrac{d\;}{d\v x}h=(0,0,1)$. These are by definition linearly dependent (by nature of it being a singularity, i.e. that $\op B_1=0$), but to use \cref{def:sings}, any pair of these gradients must remain linearly independent, which we see fails if $k=0$. Although the cusp obtained if $k=0$ would not be considered degenerate in standard bifurcation theory, this prevents the choice of $\op B_2$, over the alternatives $\op B_{2,i}$ for $i=1,2,3,$ being unique. However, this also signals here that the variable $z$ appears in a trivial fashion, and the resolution then is simply to isolate off the $z$ dependence, and study only the planar system $(f,g)=(y+x^3+\alpha x+\beta,\;y)$, for which one will find that the cusp is full and can again be found using \cref{def:sings}.


\subsection{A cusp with a degeneracy}\label{sec:cusptie}

Consider the planar vector field
\begin{align}
(f,g)&=\bb{y^2+x^3-\alpha x^2+\gamma x-\beta,\;y-k}\;,
\end{align}
with critical points $(x_*,y_*)=(x_*,k)$, where $x_*$ are the roots of $x_*^3-\alpha x_*^2+k x_*-\beta+\gamma^2=0$. Let us take $(\alpha,\beta)$ as bifurcation parameters, and so evaluate
\begin{subequations}
\begin{align}
\op B_1{}_*&=\abs{\sfrac{\partial(f,g)}{\partial{(x,y)}}}_*=\dabss{\begin{array}{cc}\gamma-2\alpha x+3x^2&2y\\0&1\end{array}}_*=\gamma-2\alpha x_*+3x_*^2\;,\nonumber\\
\op B_2{}_*&=\abs{\sfrac{\partial(\op B_1,g)}{\partial{(x,y)}}}_*=\dabss{\begin{array}{cc}6x-2\alpha&0\\0&1\end{array}}_*=6x_*-2\alpha\;,\\
\op G_{2,1}^{\alpha\beta}{}_*&=\abs{\sfrac{\partial(f,g,\op B_1,\op B_{2,1})}{\partial{(x,y,\alpha,\beta)}}}_*=-4\alpha\;,\nonumber\\
\op G_{2,2}^{\alpha\beta}{}_*&=\abs{\sfrac{\partial(f,g,\op B_1,\op B_{2,2})}{\partial{(x,y,\alpha,\beta)}}}_*=8\alpha k\;.
\end{align}
\end{subequations}
There are folds where $\op B_1{}_*=0$ when $(\beta+\sfrac2{27}a^3-k^2-\sfrac13\alpha\gamma)^2=4(\alpha^2-3\gamma)^3$ for which $x_*=\sfrac13\alpha\pm\sfrac13\sqrt{\alpha^2-3\gamma}$. 

There are cusps when $\alpha=\pm\sqrt{3\gamma}$, $\beta=k^2\pm(\gamma/3)^{3/2}$, for which $x_*=\pm\sqrt{\gamma/3}$, hence we see these only occur for $\gamma>0$. Since the various $\op G_{2,i}{}_*$ are non-zero (and one finds similar results considering any pair of parameters from $\alpha,\beta,\gamma$), we see that the cusps are non-degenerate away from the origin of this parameter space. Note the importance of $k\neq0$ for non-degeneracy. 

\begin{figure}[h!]\centering\includegraphics[width=0.7\textwidth]{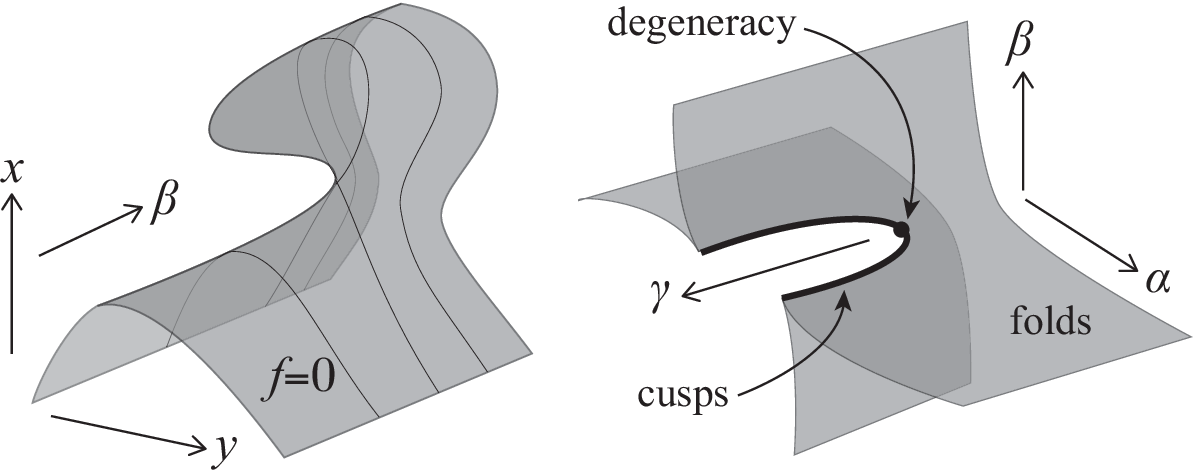}
\vspace{-0.3cm}\caption{\small\sf The hypersurface $f=0$ changing with $\beta$ (left) shown for $\alpha=-\hf$, $\gamma=-\sfrac1{10}$, and the fold set (right) in $(\alpha,\beta,\gamma)$ space, for the cusp in \cref{sec:cusptie}. 
}\label{fig:cusptie}\end{figure}

This is illustrated in \cref{fig:cusptie}, showing the hypersurface $f=0$ unfolding with $\beta$ (left), and the fold and cusp sets shown in parameter space (right). 

As \cref{fig:cusptie}(right) shows, the cusps collide when $\gamma=0$. 
This collision occurs at $\alpha=\beta=\gamma=0$. It is a degenerate cusp, but not a swallowtail, as we see if we evaluate $\op G_{3,i_1i_2i_3}{}_*=\abs{\sfrac{\partial(f,g,\op B_1,\op B_{2,i_1},\op B_{3,i_2i_3})}{\partial{(x,y,\alpha,\beta,\gamma)}}}_*=0$ for any $i_1,i_2\in\cc{1,2}$. 
A consequence of the degeneracy is that if we evaluate the condition to detect a swallowtail we find $\op B_3=6$, which cannot vanish, hence the system cannot have a swallowtail point.

\subsection{Cusp: double-quadratic form}\label{sec:cuspquad}

\begin{figure}[h!]\centering\includegraphics[width=0.47\textwidth]{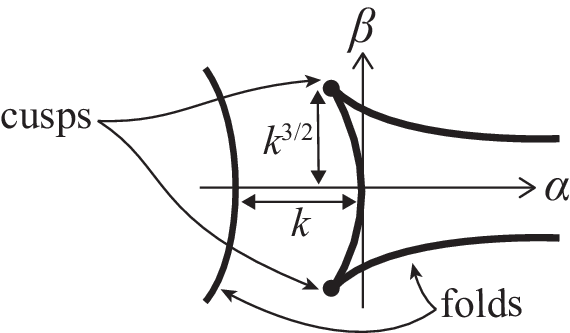}
\vspace{-0.3cm}\caption{\small\sf Bifurcation diagram of the double quadratic cusp in $(\alpha,\beta)$ parameter space, which becomes degenerate when $k=0$.  
}\label{fig:cuspdble}\end{figure}

A less obvious form for the cusp is 
\begin{align}
(f,g)&=\bb{x^2-y-\alpha,\;y^2-\beta x-ky}\;,
\end{align}
with critical points $(x_*,y_*)$ that are roots of $x_*^2-y_*-\alpha=y_*^2-\beta x_*-ky_*=0$. For the \BG conditions we have
\begin{subequations}
\begin{align}
\op B_1{}_*&=\abs{\sfrac{\partial(f,g)}{\partial{(x,y)}}}_*=\dabss{\begin{array}{cc}2x&-1\\-\beta&2y-k\end{array}}_*=4x_*y_*-2kx_*-\beta\;,\nonumber\\
\op B_2{}_*&=\abs{\sfrac{\partial(\op B_1,g)}{\partial{(x,y)}}}_*=\dabss{\begin{array}{cc}4y-2k&4x\\-\beta&2y-k\end{array}}=2(2y_*-k)^2+4\beta x_*\;,\\
\op G_{2,1}{}_*&=\abs{\sfrac{\partial(f,g,\op B_1,\op B_{2,1})}{\partial{(x,y,\alpha,\beta)}}}_*=\dabss{\begin{array}{cccc}2x&-1&-1&0\\-\beta*2y-k&0&-x\\4y-2k&4x&0&-1\\4\beta&16y-8k&0&4x\end{array}}_*\nonumber\\&=-12(k-2y_*)(\beta-2kx_*+4x_*y_*)\;,\nonumber\\
\op G_{2,2}{}_*&=\abs{\sfrac{\partial(f,g,\op B_1,\op B_{2,2})}{\partial{(x,y,\alpha,\beta)}}}_*=\dabss{\begin{array}{cccc}2x&-1&-1&0\\-\beta*2y-k&0&-x\\4y-2k&4x&0&-1\\16x&4&0&0\end{array}}_*\nonumber\\&=4(\beta-2x_*(3k+8x_*^2-6y*))\;.
\end{align}
\end{subequations}
There are folds where $4x_*y_*-2kx_*=\beta$. While it is not easy to express them explicitly, we can solve the system $f=g=\op B_1=0$ for $(x,\alpha,\beta)$, to find that they are given parametrically by 
\begin{align}
\bb{x(s),y(s)}&=\big(\pm\sqrt{\sfrac{s(s-k)}{2(2s-k)}},\;s\big)\;,\nonumber\\
\bb{\alpha(s),\beta(s)}&=\big(\sfrac{s(k-3s)}{2(2s-k)},\;\pm\sqrt{s(s-k)(2s-k)}\;\big)\;,
\end{align}
as illustrated in \cref{fig:cuspdble}, clearly existing only for $s(s-k)(2s-k)>0$. 

We have not shown here that the folds are non-degenerate, since we are most interested here in finding the cusps. However, assuming $k\neq0$, one finds for a fold with respect to $\alpha$ that $\op G^\alpha_1=\ord{k^2}$, so the folds are full. For a fold with respect to $\beta$, however, we find $\op G^\beta_1=\ord{\sqrt s}$, so the folds are full except at $s=0$, corresponding to the points $\alpha=0$ and $\alpha=-k$ on $\beta=0$ in \cref{fig:cuspdble}; here we have a situation similar to \cref{sec:bogdanovtakens} where a $\beta$-parameter path bounces off the fold curve. 

There are a pair of cusps which lie at $\alpha=(k\sqrt3-2k)/4$, $\beta=\pm k^{3/2}/3^{3/4}$, at which $x_*=\mp\sqrt{k/4\sqrt3}$, $y_*=\hf(1-\sfrac1{\sqrt3})k$. 
At the cusps we have $\op G_{2,1}{}_*=\mp8k^{5/2}/3^{1/4}$, $\op G_{2,1}{}_*=\pm3^{1/4}8k^{3/2}$, so these are full for $k\neq0$. We see throughout here that $k\neq0$ is essential here for non-degeneracy.



{
\bibliography{HiCatsRsim.bbl}
\bibliographystyle{plain} 
}

\end{document}